\documentclass[graybox]{svmult}

\usepackage{type1cm}        % activate if the above 3 fonts are
\usepackage{makeidx}         % allows index generation
\usepackage{graphicx}        % standard LaTeX graphics tool
\usepackage{multicol}        % used for the two-column index
\usepackage[bottom]{footmisc}% places footnotes at page bottom
\usepackage{bm}
\usepackage{standalone}
\usepackage[normalem]{ulem}
\usepackage{lipsum}
\usepackage{float}
\usepackage{graphicx}
\usepackage{algorithm}
\usepackage{amssymb, amsmath, amsfonts}
\usepackage[caption=false]{subfig} % <- Preamble
\usepackage{pgfplots} % <- Preamble
\usepackage{xcolor}
\usepackage{mathtools}
\usepackage{bm}
\usepackage{newtxtext}  % 
\usepackage[varvw]{newtxmath}

\DeclareMathOperator*{\argmax}{arg\,max}

\makeatletter
\def\bbordermatrix#1{\begingroup \m@th \@tempdima 4.75\p@ \setbox\z@\vbox{%
    \def\cr{\crcr\noalign{\kern2\p@\global\let\cr\endline}}%
    \ialign{$##$\hfil\kern2\p@\kern\@tempdima&\thinspace\hfil$##$\hfil &&\quad\hfil$##$\hfil\crcr
      \omit\strut\hfil\crcr\noalign{\kern-\baselineskip}%
      #1\crcr\omit\strut\cr}}%
  \setbox\tw@\vbox{\unvcopy\z@\global\setbox\@ne\lastbox}%
  \setbox\tw@\hbox{\unhbox\@ne\unskip\global\setbox\@ne\lastbox}%
  \setbox\tw@\hbox{$\kern\wd\@ne\kern-\@tempdima\left[\kern-\wd\@ne \global\setbox\@ne\vbox{\box\@ne\kern2\p@}%
    \vcenter{\kern-\ht\@ne\unvbox\z@\kern-\baselineskip}\,\right]$}%
  \null\;\vbox{\kern\ht\@ne\box\tw@}\endgroup}
\makeatother
\DeclareMathAlphabet{\mathdutchcal}{U}{dutchcal}{m}{n} \SetMathAlphabet{\mathdutchcal}{bold}{U}{dutchcal}{b}{n}
\DeclareMathAlphabet{\mathdutchbcal}{U}{dutchcal}{b}{n} \newcommand{\at}[2][]{#1|_{#2}}

\makeindex             % used for the subject index
\begin{document}
%\title*{Foraging Scout Efficiency in \emph{Aphaenogaster senilis} with Different Personalities on the Honeycomb Lattice: First-Passage Statistics with Heterogeneous Walkers with and without Persistence}

%\title*{Foraging Scout Efficiency in \emph{Aphaenogaster senilis} with Different Personalities on the Honeycomb
%Lattice: First-Passage Statistics with Heterogeneous Walkers with and without Persistence}

%FEDE: I propose a more general title. (also note that the previous was wrong in that foraging is indeed a combination
%of scouts and recruits...)
\title*{Collective Foraging and Behavioural Syndromes in Ants: First-Passage Statistics with Heterogeneous Walkers on a Honeycomb Lattice}

\titlerunning{Application of persistent random walk theory to \textit{A. senilis} foraging} \author{Daniel
Marris$^*$\orcidID{0000-0002-1255-3126}, \\ Pol Fern{\'a}ndez-L{\'o}pez$^*$\orcidID{0000-0003-0955-5945}, \\ Frederic
Bartumeus\orcidID{0000-0001-6908-3797} and \\  Luca Giuggioli\orcidID{0000-0002-2325-4682}} \authorrunning{D. Marris and
P. Fern{\'a}ndez L{\'o}pez {\em et al.}} \institute{ Daniel Marris$^*$ \at School of Engineering Mathematics and Technology,
University of Bristol, BS8 1TW, UK. %\email{daniel.marris@bristol.ac.uk}
\and 
Pol Fern{\'a}ndez-L{\'o}pez$^*$ \at Theoretical and Computational Ecology, Centre for Advanced Studies of Blanes, C/
Accés a la Cala Sant Francesc 14, 17300, Girona, Spain. %\email{pfernandez@ceab.csic.es}
\and 
Frederic Bartumeus \at Theoretical and Computational Ecology, Centre for Advanced Studies of Blanes, C/ Accés a la Cala
Sant Francesc 14, 17300, Girona, Spain.  \\
Centre for Research on Ecology and Forestry Applications (CREAF),  08193 Barcelona, Spain. \\
Catalan Institution for Research and Advanced Studies (ICREA), Passeig de Lluís Companys, 23, 08010 Barcelona, Spain.
%\email{fbartu@ceab.csic.es}
\and
Luca Giuggioli \at School of Engineering Mathematics and Technology, University of Bristol, BS8 1TW, UK. \\
Corresponding author: \email{luca.giuggioli@bristol.ac.uk}
\and
}
%
% Use the package "url.sty" to avoid problems with special characters used in your e-mail or web address
%

\maketitle
\def\thefootnote{*}\footnotetext{These authors contributed equally to this work} \abstract{Behavioral heterogeneities in animals, also known as syndromes, play a crucial role in understanding how natural populations flexibly adapt to environmental changes. In ant species
like \textit{Aphaenogaster senilis}, two key roles in collective foraging are commonly recognised: scouts, who discover
food patches, and recruits, who exploit these patches and transport food back to the nest. These roles involve distinct
movement patterns and exploratory behaviours. In this chapter, we develop a correlated random walk model on a bounded
honeycomb lattice to interpret and replicate empirical observations of foraging ants in an enclosed arena with honeycomb
tiling. We do so by extending the theory of first-passage processes for $\mathcal{N}$ random walkers when individuals
belong to a heterogeneous population. We apply this theory to examine how individual behavioural heterogeneity in ants
affects collective foraging efficiency, focusing on first-passage time statistics for nest-to-patch and patch-to-patch
movements. With the combined use of the mathematical model and the controlled experimental setup we evaluate (i) the
impact of distinct movement strategies by scouts and recruits on finding food items and (ii) whether ants practice
strict central place foraging or utilise previously discovered patches as starting points for further exploration.}

\section{Introduction}\label{sec: intro} Social groups are rarely uniform or cohesive units; instead, individuals within
these groups often exhibit consistent behavioural differences, commonly referred to as ``personalities''
\cite{Jolles2020}. Understanding individual heterogeneity is essential for grasping the variability in life histories
and population dynamics \cite{chesson1991}. Moreover, it significantly influences the functionality of key ecological
responses. In eusocial species, such as ants and termites, individual heterogeneity is exploited to associate different
roles to different tasks \cite{antecologybook}, e.g., typically one female and several males are reproductively active,
while nonbreeding individuals care for the young, protect the colony, and provide resources. A key debate in this regard
is whether these roles represent specialised task forces designed to address different needs or if individuals switch
between tasks to adapt to inherent environmental uncertainties \cite{Gordon2016}.

In the context of foraging processes, ants are well known to have evolved various strategies for efficient localisation
and exploitation of food resources \cite{antecologybook} with two common strategies being group recruitment and mass
recruitment. In group recruitment, a scout discovers a food source and returns to the nest to recruit a small number of
nestmates by leaving a chemical trail that only the scout itself can trace. This strategy, where nestmates are
followers, suits smaller or scattered food sources. On the other hand, mass recruitment, where the scout lays a strong
pheromone trail that may attract many nestmates, is faster and involves more ants, ideal for large or abundant food
sources. Key determinants to select the appropriate strategy are the environmental conditions and the type of food
available. For instance, some species primarily rely on individual foraging but switch to group or mass recruitment as
food sources increase in abundance and predictability, allowing ant colonies to optimise resource acquisition across
different scenarios. The number of scouts and recruits involved in food collection is also flexible, varying dynamically
as the food is being collected and with the amount and predictability of resources. This adaptability raises intriguing
questions about the role of these distinct behaviours, how colonies adjust their numbers, and how these adjustments
impact collective success.

Experiments that accurately track animal movement are a convenient approach to learn about foraging strategies. Y-maze
laboratory arenas, where the ants are confined to the perimeters of tiled hexagons, are particularly suited to
understanding ant behaviour as they offer a reliable and practical means of quantifying decision-making
\cite{Czaczkes2018}. Here we employ the empirical set-up used in previous experiments \cite{cristin2024spatiotemporal}
to observe the foraging behaviour of ants (\textit{Aphaenogaster senilis}) within a large ($2\times 1\,\text{m}^{2}$)
honeycomb lattice arena. 

\textit{A. senilis} has been selected because of its known group recruitment foraging strategies whereby scouts and
recruits have distinct roles~\cite{campos2014,martin2024}. Recent studies on a Y-maze arena have identified significant
disparities in the movement patterns of these two types of ants \cite{pol2024liquid}, with quantifiable differences in
the probabilities of moving left, right, or backward at each step. Scouts exhibit less sinuous movement and are less
likely to move backward compared to recruits. In contrast, recruits display movement patterns more akin to a diffusive
walk, with an approximately uniform sampling of the three possible directions at each vertex of the honeycomb lattice.
Consequently, scouts travel greater distances from the nest and spend more time exploring the arena, while recruits
typically remain within a confined region, usually no farther than 25 cm from the nest. One may thus expect that the
differences in numbers and exploratory behaviours of scouts and recruits would lead to markedly different spatial
spreading and coverage patterns.

%To assess the roles of scouts and recruits in producing effective foraging it is important to have a better
%understanding of the differences in their exploratory behaviour, more specifically, the spreading and spatial coverage
%patterns achieved by the two behavioural types of ants.  

While ants are known to perform central place foraging \cite{gordon1979central}, returning to a central location (like a
nest or den) after foraging trips, it is not clear what movement strategies they employ in environments with multiple
food patches, as new patches may be discovered while previous ones are still being exploited. In these complex,
multi-target scenarios, both the nest and the exploited patches could serve as departure points for further searches by
either scouts or recruits. Here we investigate whether exploratory ants follow a strict central place foraging strategy
(nest-to-patch search) or use previously discovered patches as starting points for further exploration (patch-to-patch
search). 

To answer quantitatively these questions we develop a discrete space-time, non-Markovian, random walk model on a bounded
honeycomb lattice that faithfully represents the experimental arena. Since the honeycomb lattice is non-Bravais, to
model movement and search in the experimental arena, we utilise the theory of random walks with internal states to
capture the periodic heterogeneity in the domain sites \cite{marris2023exact, montroll1969random} and include a one-step
memory in the model via the theory of persistent random walks \cite{larralde2020first, marris2024persistent}. We use
this model first with a single walker to investigate the spreading (mean square displacement and mean number of distinct
visited sites) in the infinite lattice, quantifying how quickly scouts and recruiters explore the space compared to a
diffusive walker. We then study the so-called first first-passage time (fFPT) distribution of a population of
independent and heterogeneous walkers. We do so by extending well-known theory \cite{weiss1983order} to account for a
population of distinct walkers, that is with unequal movement statistics. We further generalise the theory to consider
multiple targets and calculate the time-dependent splitting probability, defined as the probability as a function time
that one member of the population reaches specific target(s) before any other individual has reached any other target.
Similarly to the case of one single walker  \cite{marris2024persistent, tejedor2012optimizing},  with multiple walkers
one may expect that an increase in persistence hinders search times. Using the splitting probability we show in fact
that a highly persistent population may miss close-by targets. In contrast, when one uses experimentally relevant
parameters, we show that the local target is found first.

By employing a semi-analytic procedure,  we study the ant search behaviour on an exact replica of the experimental arena
with hard reflecting boundaries. We explore the fFPT distribution for the $\mathcal{N}$ ants observed in nine different
experimental realisations, and demonstrate that in the majority of cases, the empirical measurements coincide much
closer to the mode of the fFPT distribution than its mean. Moreover, by identifying the differences in the ordering of
events in nest-to-patch and patch-to-patch search, we elucidate that the empirical timescales to find the second patch
match closer with the nest-to-patch case.
\begin{figure}
    \centering
    \includegraphics[width = \linewidth]{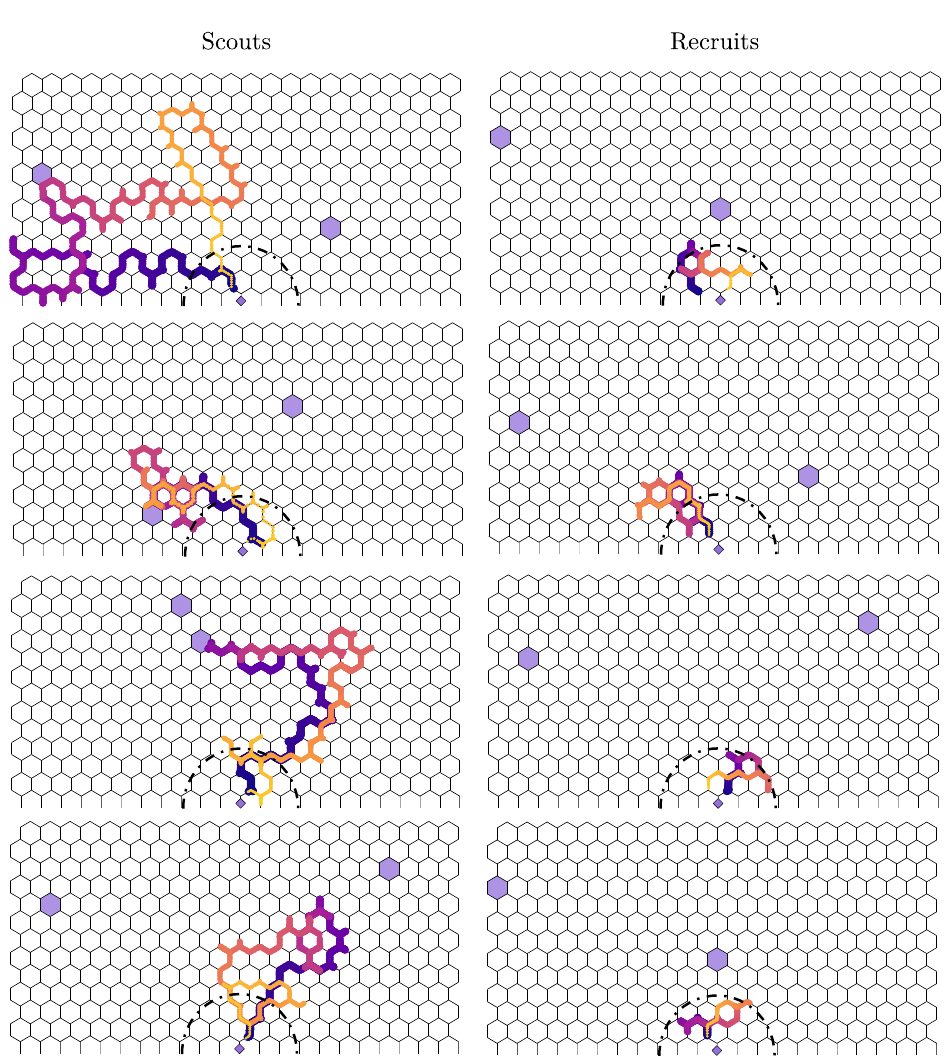}
    \caption{Subset of empirical trajectories on the Y-maze from scouts (left) and recruits (right). The purple diamond
    on the bottom centre depicts the entry point to the experimental arena ($2\times 1\,\text{m}^2$). The purple
    hexagons are the locations for the food patches, changing across different experimental trials. The colour gradient
    in the trajectories represents the temporal evolution, indicating where the trajectories start (blue) and end
    (yellow). Trajectories displaying an average distance to the nest exceeding 25 cm (dashed semicircle) were
    classified as scouts or recruits otherwise. The thickness of the zig-zag lines aids visualization of time: thick
    segments indicate early times, while thin segments correspond to late times. }
    \label{fig: indiv_tracks}
\end{figure}

The chapter is organised as follows. The details of the experimental procedures are described in Sec. \ref{sec: exp}. In
Sec. \ref{sec: me} we derive the formalism of the honeycomb correlated random walk on the unbounded (infinite) lattice
and then use it to explore coverage efficiency in Sec. \ref{sec: coverage}. We then turn our attention to the first
first-passage time of $\mathcal{N}$ walkers for which we derive a general formalism for the splitting probability with
multiple targets when the population is composed of non-identical walkers. We study some theoretical examples on bounded
lattices in Sec. \ref{sec: theor_ffpt}, while the experimentally relevant model on the Y-maze arena is worked out in
Sec. \ref{sec: exp_ffpt}. Empirical findings  follow in Sec. \ref{sec: results}, and conclusions form Sec. \ref{sec:
discussion}.

\section{Experimental protocol in the Y-maze arena}\label{sec: exp} Foraging experiments were carried out with
\textit{Aphaenogaster senilis}, a Mediterranean ant species displaying flexible foraging strategies, ranging from
individual to group recruitment \cite{Cerda2009}. To reproduce the natural spatiotemporal scales at which these ants
forage, we recorded three hours of video per trial, capturing their movements within a $2\times 1\,$ meters squared
honeycomb-like arena  (see Figure 1 in ref. \cite{cristin2024spatiotemporal}). This Y-maze pattern is widely utilized in
studies of animal decision-making (e.g., \cite{Czaczkes2018}), and simplifies the quantification of animal behaviour and
movement since at each vertex the walker has only three choices: to go left, right or backwards. Additionally, spatial
discreteness of the arena lends itself to the use of the present random walk theory, and other statistical
physics frameworks such as spin glasses~\cite{cristin2024spatiotemporal,Cristin2020}. 

The main purpose of the experiments was to test how ant foraging dynamics adapted to highly unpredictable environments.
Accordingly, food sources were randomly distributed through the experimental arena, changing food placement at each
experimental trial (nine experiments with different food placements). The food resources consisted of two
\textit{Tenebrio molitor} larvae cut into six pieces arranged in the vertices of two hexagons of the arena, signifying
food patches.

We tracked individual ant trajectories from the videos at a temporal resolution of 2 Hz. We analysed several aspects of
their movement and classified them into scouts or recruits according to the tracks' average distance to the nest. After,
the tracks were discretised in time at 1-second intervals, each position was assigned to the closest node (Y-shaped intersection) in the lattice.
We then estimated the turning probabilities, computing whether the ant remained in the same node
($c$), or turned to the right ($r$), left ($l$), or backward ($b$), relative to its last position. This was performed
separately for scouts and recruits. The relevant parameters, averaged over all experiments, for the recruits are $r\approx0.0818$, $b\approx0.0539$,
$l\approx0.0821$, and $c\approx0.7822$, while for the scouts we found $ r\approx 0.1039$, $b\approx 0.0358$ and $l
\approx 0.1008$, and $c\approx0.7595$, with all movement parameters rounded to 4 decimal places. In
Figure~\ref{fig: indiv_tracks} we represent some example tracks for both scouts and recruits, to showcase their
different movement patterns.

\section{Master Equation for a correlated walk in a honeycomb lattice in unbounded space}\label{sec: me}
To model the
correlated movement of the ants on the non-Bravais honeycomb lattice we use the theory of lattice random walk with
internal states \cite{montroll1969random,weiss1994aspects}. The model requires six internal states, representing both
the physical site types (spatial heterogeneity) \cite{marris2023exact,zumofen1982energy} and the direction last
travelled (one-step memory) \cite{larralde2020first, marris2024persistent}. For the spatial degree of freedom we follow
Montroll's procedure of embedding the honeycomb lattice into the $\mathbb{Z}^2$ square lattice
\cite{montroll1969random}, where each unit cell contains two site types ($\triangleright$ and $\triangleleft$), referred
to henceforth as positive $(\mathfrak{p})$ and negative $(\mathfrak{n})$ sites. As described in Sec. \ref{sec: exp}, the
movement parameters of the model represent the probability of making a right turn $(r)$ a left turn $(l)$ or a backtrack
$(b)$. Since the direction previously travelled must be known to determine in which direction the walker will turn, each
physical site is then decomposed into three internal states representing the movement direction last travelled. We also
allow the walker at each time step to remain on a site with probability $c$, such that $l+r+b+c=1$. This construction
allows to express the non-Markov movement steps on the honeycomb lattice as a higher-order Markov chain
\cite{van1998remarks}. 

\begin{figure}[h]
    \includegraphics*[width=\textwidth]{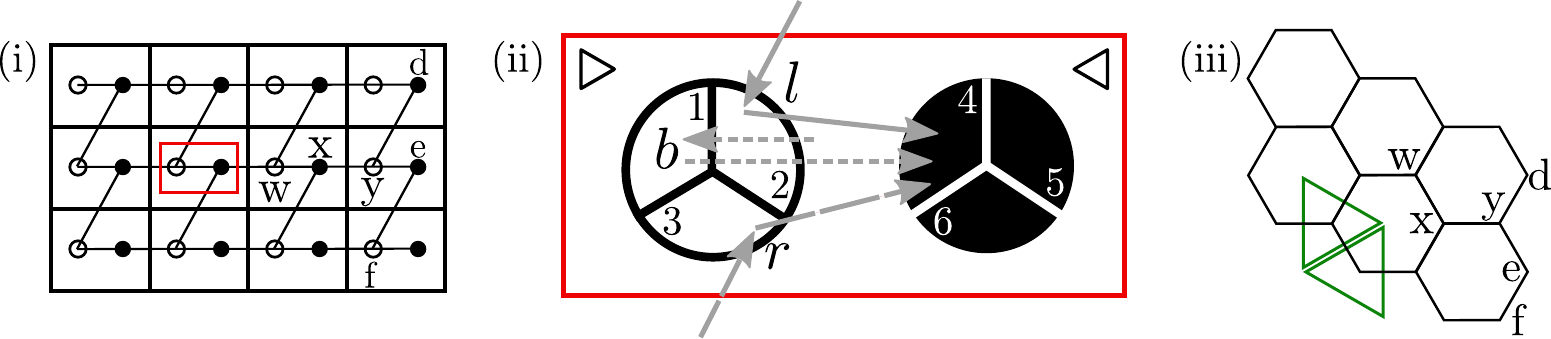}
    \caption{(Colour online). A schematic showing the theoretical domain of interest. (i) The Montroll
    \cite{montroll1969random} embedding of the honeycomb lattice into the $\mathbb{Z}^2$ lattice, whereby a visual
    comparison between (i) and (iii) confirms that one requires one more unit cell along each dimension in the integer
    lattice (i) than the number of hexagons in that direction in the hexagonal lattice (iii). Panel (ii) details the
    decomposition of the positive ($\triangleright$) and negative ($\triangleleft$) sites into their memory components
    and represents a `blow-up' of the red box in plot (i). We show, via the arrows, the possibilities to enter state 4
    (with sojourns omitted for visual clarity). These are taking a left turn from state 2 (having previously lowered the
    $n_2$ coordinate) a right turn from state 3 (having previously raised the $n_2$ coordinate) or making a backtrack
    from state 1 (having previously moved from negative to positive site in the unit cell). Subfigures (i) and (iii) are
    in the style of Fig. 2  of ref. \cite{montroll1969random}. }
    \label{fig: lattice_schematic}
\end{figure}

More specifically, we describe the one step non-Markov dynamics via the coupled Master equation
\begin{align}
    P(\!\bm{n}, \!1, t\!+\!1\!) &\!=\! b P(n_1, n_2, 4, t) \!+\! l P(n_1, n_2, 5, t) \!+\! r P(n_1, n_2, 6, t) \!+\! c P(\bm{n}, 1, t),\nonumber \\ 
    P(\!\bm{n},\! 2, t\!+\!1\!) &\!=\! r P(n_1, n_2 \!+\! 1, 4, t) \!+\! b P(n_1, n_2 \!+\! 1, 5, t) \!+\! l P(n_1, n_2\!+\!1, 6, t) \!+\! c P(\bm{n}, 2, t), \nonumber \\ 
    P(\!\bm{n}, \!3, t\!+\!1\!) &\!=\! l P(n_1 \!-\! 1, n_2, 4, t) \!+\! r P(n_1 \!-\! 1, n_2, 5, t) \!+\! b P(n_1 \!-\! 1, n_2, 6, t) \!+\! c P(\bm{n}, 3, t), \nonumber\\ 
    P(\!\bm{n}, \!4, t\!+\!1\!) &\!=\! b P(n_1, n_2, 1, t) \!+\! l P(n_1, n_2, 2, t) \!+\! r P(n_1, n_2, 3, t) \!+\! c P(\bm{n}, 4, t),\nonumber \\ 
    P(\!\bm{n},\! 5, t\!+\!1\!) &\!=\! r P(n_1, n_2 \!-\! 1, 1, t) \!+\! b P(n_1, n_2 \!-\! 1, 2, t) \!+\! l P(n_1, n_2\!-\!1, 3, t) \!+\! c P(\bm{n}, 5, t), \nonumber \\ 
    P(\!\bm{n}, \!6, t\!+\!1\!) &\!=\! l P(n_1\!+\!1, n_2, 1, t) \!+\! r P(n_1\!+\!1, n_2, 2, t) \!+\! b P(n_1\!+\!1, n_2, 3, t) \!+\! c P(\bm{n}, 6, t), 
    \label{eq: u_ME}
\end{align}
where the notation $P(\bm{n}, m, t)=P(n_1, n_2, m, t)$ denotes the probability at time $t$ of occupying internal state
$m$ at site $\bm{n}=(n_1, n_2)$ and each timestep represents a one second interval. States $m \in \{1, 3\}$ represent
the movement directions into positive states, while states $m \in \{4, 6\}$ represent the movement directions into
negative states (see Fig. \ref{fig: lattice_schematic} for pictorial representation).

Equation (\ref{eq: u_ME}) is completely general, modelling a wide range of movement behaviour. The diffusive limiting
case treated in \cite{marris2023exact,montroll1969random,zumofen1982energy} can be recovered by taking $r = l = b = q/3$
($0< q\leq 1$). In contrast, if $r = 1$ or $l = 1$, the motion is deterministic whereby the walker traverses the
perimeter of one hexagon in a prescribed direction for all times. 

To seek the solution of Eq. (\ref{eq: u_ME}), it is more convenient to re-write the coupled scalar equation in matrix
form as 
\begin{eqnarray}
    \bm{P}(n_1, n_2, t+1) &= \mathbb{A}\cdot \bm{P}(n_1 - 1, n_2, t) + \mathbb{B}\cdot \bm{P}(n_1 + 1, n_2, t) +
    \mathbb{C}\cdot \bm{P}(n_1, n_2-1, t) \nonumber \\ &+ \mathbb{D}\cdot \bm{P}(n_1, n_2+1, t)+\mathbb{E}\cdot
    \bm{P}(n_1, n_2, t),
    \label{eq: vec_ME}
\end{eqnarray}
where here, and elsewhere, capitalised bold symbols denote column vectors and where the matrices containing the
transition probabilities, viz. $\mathbb{A}, \mathbb{B}..., \mathbb{E}$, are defined in Appendix 1. We introduce the
initial condition $\bm{P}(n_1, n_2, 0) = \delta_{\bm{n}, \bm{n_0}}\bm{U}_{m_0}$, which states the walkers are intitally
localised at some site $\bm{n}$, with the probability of beginning their trajectory in some internal state $m$
distributed by $\bm{U}_{m_0}$. In our case, the elements of $\bm{U}_{m_0}$ are dependent on whether the walker starts in
a positive or negative state as well as in which direction the walker is assumed to have entered that site from. With
little information on the direction travelled at $t=0$ within the nest, we assume the walker entered via a right, left
or back turn with equal probability leading to $\bm{U}_{\mathfrak{p}} = [1/3, 1/3, 1/3, 0, 0, 0]^{\intercal}$ if the
walker started in a positive state and $\bm{U}_{\mathfrak{n}} = [ 0, 0, 0, 1/3, 1/3, 1/3]^{\intercal}$ otherwise. 

By performing the Fourier transform, $\widehat{f}(\xi)=\sum_{n\in \mathbb{Z}}\text{e}^{i \xi n}f(n)$, and the
$z$-transform, $\widetilde{f}(z) = \sum_{t=0}^{\infty}z^t f(t)$, on Eq. (\ref{eq: vec_ME}) we find 
\begin{equation}
    \widehat{\widetilde{\bm{P}}}_{\bm{n_0}, m_0}(\xi_1, \xi_2, z) = \text{e}^{-i\left[\xi_1 n_{01} + \xi_2 n_{02} \right]}\left[\mathbb{I}-z\bm{\lambda}(\xi_1, \xi_2)\right]^{-1}\cdot \bm{U}_{m_0},
    \label{eq: corr_FZ}
\end{equation}
where $\bm{\xi}=(\xi_1, \xi_2)$ are the Fourier components along each Cartesian axis and
\begin{equation}
    \bm{\lambda}(\xi_1, \xi_2) = 
    \bbordermatrix{      & \nwarrow & \swarrow & \longrightarrow & \searrow & \nearrow & \longleftarrow \cr
    \nwarrow & c & 0 & 0 & b & l & r \cr
    \swarrow & 0 & c & 0 & r \text{e}^{-i \xi_2} & b\text{e}^{-i \xi_2} & l \text{e}^{-i \xi_2} \cr
    \longrightarrow & 0 & 0 & c & l \text{e}^{i \xi_1} & r \text{e}^{i \xi_1} & b \text{e}^{i \xi_1} \cr
    \searrow & b & l & r & c & 0 & 0 \cr
    \nearrow & r \text{e}^{i \xi_2} & b \text{e}^{i \xi_2} & l \text{e}^{i \xi_2} & 0 & c & 0 \cr
    \longleftarrow & l \text{e}^{-i \xi_1} & r \text{e}^{-i \xi_1} & b \text{e}^{-i \xi_1} & 0 & 0 & c    
    }.
    \label{eq: sf}
\end{equation}
The arrows around the matrix $\bm{\lambda}(\xi_1, \xi_2)$ represent the movement on the honeycomb lattice encoded by the
element in the matrix where $\bm{\lambda}(\xi_1, \xi_2)_{i,j}$ governs the movement from element $j$ to element $i$. To
illustrate, consider the element $\bm{\lambda}(\xi_1, \xi_2)_{5, 2}$. To move from state 2 to state 5, one must hop from
a positive state to the negative state in the unit cell above via a backtracking from the previous movement direction.
This corresponds to a movement in the South-West direction followed by a movement back in the North-East direction as
may be evinced by the arrows surrounding Eq. (\ref{eq: sf}) and the schematic in Fig. \ref{fig: lattice_schematic}. We
note that the choice of the `positive' Fourier transform, $\widehat{f}(\xi)=\sum_{n\in \mathbb{Z}}\text{e}^{i \xi
n}f(n)$, leads to exponents in the structure function, which reflects the direction of movement. For example, an element
$p\text{e}^{i \xi_1}$ represents the possibility of a step to the right with probability $p$ \cite{klafter2011first}. An
alternative definition of the Fourier transform with negative exponent may be taken instead \cite{LucaPRX}, which leads
to an equivalent representation of the dynamics as Eq. (\ref{eq: corr_FZ}), whereby $(\xi_1, \xi_2)\rightarrow (-\xi_1,
-\xi_2)$. 

From Eq. (\ref{eq: corr_FZ}), we take the inverse Fourier and $z$-transforms to find the unbounded propagator or lattice
Green's function 
\begin{equation}
    \bm{P}_{\bm{n_0}, m_0}(n_1, n_2, t)\! =\! \frac{1}{4\pi^2}\int_{-\pi}^{\pi}\int_{-\pi}^{\pi}\!\text{e}^{-i\left[\xi_1(n_1 - n_{01}) + \xi_2(n_2 - n_{02}) \right]}\bm{\lambda}(\xi_1, \xi_2)^t\cdot \bm{U}_{m_0}\text{d}\xi_1 \text{d}\xi_2,
    \label{eq: corr_t}
\end{equation}
which represents a $6\times 1$ column vector. The probability of occupying the positive state in $(n_1, n_2)$ is
$P_{\bm{n}_0, m_0}^{(\mathfrak{p})}(n_1, n_2, t) = \sum_{i=1}^{3}\bm{P}_{\bm{n}_0,m_0}(n_1, n_2, t)_i$, equivalently
$P_{\bm{n}_0, m_0}^{(\mathfrak{p})}(n_1, n_2, t) = \sum_{i=1}^{3}P_{\bm{n}_0,m_0}(n_1, n_2, i, t)$, while for the
corresponding negative state one takes $P_{\bm{n}_0,m_0}^{(\mathfrak{n})}(n_1, n_2, t)
=\sum_{i=4}^{6}\bm{P}_{\bm{n}_0,m_0}(n_1, n_2,t)_i$. 

\section{Coverage Metrics}\label{sec: coverage} There are two fundamental quantities to elucidate how efficiently a
random walker explores the space over time: the mean-squared displacement (MSD), $\Delta\bm{n}(t)$, and the mean number
of distinct visited sites (MNDVS) up to time $t$, $S(t)$. We study here both quantities for scouts and recruits on the
unbounded lattice. 

\subsection{Mean Squared Displacement}
We begin by defining displacement on the honeycomb lattice. Although other definitions have been proposed
\cite{wilson2019displacement}, the most natural definition of displacement on networks is simply the shortest path
\cite{gallos2004random}. Here, as we embed the honeycomb lattice into the square lattice, we define displacement as the
shortest path on the square lattice. While this definition provides an underestimation of the shortest path on the
honeycomb lattice, it allows for exact calculations. 

To calculate the MSD, we begin by assuming without loss of generality that the walker starts in the positive site,
$P(\bm{n}, 0) = \delta_{n_1, 0}\delta_{n_2, 0}\delta_{m_0, \mathfrak{p}}$. Using this initial condition in Eq. (\ref{eq:
corr_FZ}) leads to 
\begin{equation}
    \widehat{\widetilde{\bm{P}}}_{\bm{n_0}, \mathfrak{p}}(\xi_1, \xi_2, z) = \left[\mathbb{I}-z\bm{\lambda}(\xi_1, \xi_2)\right]^{-1}\cdot \bm{U}_{\mathfrak{p}}.
    \label{eq: FZ_0IC}
\end{equation}
From Eq. (\ref{eq: FZ_0IC}), we find the MSD via the second derivative of the propagator in Fourier space as 
\begin{equation}
    \Delta\bm{n}(t) = -\mathcal{Z}^{-1}\left\{\sum_{i =1}^{2}\frac{\partial^2}{\partial \xi_i ^2 } \widehat{\widetilde{P}}^{\mathcal{(\mathfrak{p})}}_{\bm{n_0}, \mathfrak{p}}(\xi_1, \xi_2, z) + \sum_{i =1}^{2}\frac{\partial^2}{\partial \xi_i ^2 } \widehat{\widetilde{P}}^{\mathcal{(\mathfrak{n})}}_{\bm{n_0}, \mathfrak{p}}(\xi_1, \xi_2, z) \right\}\Bigg|_{(\xi_1, \xi_2)=(0,0)},
\end{equation} 
where $\mathcal{Z}^{-1}$ denotes the inverse $z$-transform. After some algebra we find
\begin{equation}
   \widetilde{\Delta\bm{n}}(z)= \frac{2 (c - 1) z + 2 (c - 1) (b - 3 c) z^2 + 
   \delta z^3 + 
   \gamma z^4}{3 (z - 1)^2 \left(z (2 c - 1) - 1\right) \left(1 + \alpha z + \beta z^2\right)},
   \label{eq: MSD_z}
\end{equation}
where 
\begin{align}
    &\alpha = 2 - 4 c - 3 (l + r), \quad \beta = 1 + 4 c^2 - c \left[4 - 6 (l + r)\right] + 3 \left[l^2 + (r - 1) (l + r)\right], \nonumber \\
    &\delta = 2 (c - 1) \left[4 c^2 - (l - 1)^2 - (l - 2) r - r^2\right], \\ 
    &\gamma = 2 (c \! -\! 1) \left[3 r \! -\! 1 \! -\! 4 c^2 (c\!+\! b) + 
    4 c \left[(l \! -\! 1)^2 + (l \! -\! 2) r + r^2\right] - 3 (r^2 \! -\! l (c \!+\! b))\right]. \nonumber
\end{align}
While the complex poles of Eq. (\ref{eq: MSD_z}) may be found analytically and the time-dependent MSD extracted, its
expression is very cumbersome. We refrain from presenting it, and instead we show analytically some limiting cases and
display the experimentally relevant cases in Fig. \ref{fig: spreading}(a). 

We begin with the diffusive limit $r = l = b = (1-c)/3$, whose time dependent MSD is (see Appendix 2) 
\begin{equation}
    \Delta \bm{n}(t) =  9^{-1}\left[4(1-c)t + 1 - (2c-1)^t\right].
    \label{eq: diff_MSD}
\end{equation}
The first term of Eq. (\ref{eq: diff_MSD}) displays the linear MSD with the dependence on the sojourn probability $c$
that one expects from diffusive motion. The second time-dependent term, which might be surprising on first inspection,
arises because at each time step one of the three movement options keeps the walker in the same unit cell. Therefore,
after one step the MSD over the embedded lattice (with $c=0$) is $\Delta \bm{n}(t) = 2/3$ compared to $\Delta \bm{n}(t)
= 1$ over the real lattice.   

Some other limiting cases include the backtracking limit ($b=1$) and the circular limit ($r=1$ or $l=1$). In the former
case, we find $\Delta \bm{n}(t) = 3^{-1}\left[1+(-1)^t\right]$, which represents a walker which hops between two sites
for all time. In the latter case, the hexagonal structure of the honeycomb lattice leads to circular-like motion,
whereby the walker traverses the perimeter of one hexagon for all time. Hence, the MSD is a piecewise linear function,
$\Delta \bm{n}(t) = 9^{-1}\left[9-(-1)^t-8\cos\left(\frac{\pi t}{3}\right)\right]$ which periodically returns to zero
with period six. In contrast, in the so-called directed $r=l=1/2$ case we find $\Delta \bm{n}(t) =9^{-1} \left[ 2^{(4 -
t)}-(-1)^t + 3(4 t - 5)\right]$, with super-linear dependence at short times before diffusive spread quickly takes over.

\subsection{Mean Number of Distinct Visited Sites}
An alternate measure of exploration efficiency is the MNDVS, which has attracted lots of attention in the diffusive
limit and has been studied on lattices with different numbers of neighbouring sites, the so-called coordination number
\cite{montroll1969random, zumofen1982energy}. Somewhat surprisingly, this quantity has been studied to a much lesser
extent when the movement is persistent with some results for the square lattice appearing only recently
\cite{larralde2020first}. 

Presently, we define a site as visited if any of its internal states have been occupied. More precisely, if for example,
state $1$ of site $\bm{n}$ is visited, then the positive site is counted and any visits to states $2$ or $3$ will not
increase the count. On the other hand, if state $4$ is visited then the negative site in $\bm{n}$ is counted. As such,
unlike the MSD, the MNDVS is unaffected by the inclusion of the spatial internal degrees of freedom.

It is well-known that the MNDVS is related to the first-passage probability via a summation of the first-passage over
all states up to some time $t$ \cite{biroli2022number,rws_on_latticesII}, which in our case is written as
\begin{equation}
    \mathcal{M}(t) = \sum_{\bm{n}}\sum_{j\in\{\mathfrak{p}, \mathfrak{n}\}}\sum_{t' = 1}^{t}F_{\bm{n_0},m_0}^{(j)}(\bm{n}, t').
    \label{eq: MNDVS_t}
\end{equation}
Via the celebrated renewal equation, the first-passage probability is known to be related to a simpler quantity, the
occupation probability \cite{rws_on_latticesII, redner2001guide}. In the case of the correlated random walk, if the
condition $\widetilde{P}_{\bm{n}, \mathfrak{n}}^{(\mathfrak{n})}(\bm{n}, z) = \widetilde{P}_{\bm{0},
\mathfrak{p}}^{(\mathfrak{p})}(\bm{0}, z)$ is met then the renewal equation reduces to the Markovian case
\cite{larralde2020first, marris2024persistent}. Here this condition is satisfied allowing us to express the
first-passage to a positive site as $ \widetilde{F}_{\bm{n_0}, m_0}^{(\mathfrak{p})}(\bm{n}, z)
=\widetilde{P}_{\bm{n_0},m_0}^{(\mathfrak{p})}(\bm{n}, z)/\widetilde{P}_{\bm{n}, \mathfrak{p}}^{(\mathfrak{p})}(\bm{n},
z) $. Passing this relation into the generating function of Eq. (\ref{eq: MNDVS_t}) and again making the harmless
assumption that the walker started in a positive state, we find 
\begin{equation}
    \widetilde{\mathcal{M}}(z) = \frac{1}{1-z}\sum_{\bm{n}}\left[\frac{\widetilde{P}_{\bm{n_0}, \mathfrak{p}}^{(\mathfrak{p})}(\bm{n}, z)}{\widetilde{P}_{\bm{n}, \mathfrak{p}}^{(\mathfrak{p})}(\bm{n}, z)}+\frac{\widetilde{P}_{\bm{n_0}, \mathfrak{p}}^{(\mathfrak{n})}(\bm{n}, z)}{\widetilde{P}_{\bm{n}, \mathfrak{n}}^{(\mathfrak{n})}(\bm{n}, z)}\right],
    \label{eq: MNDVS_z}
\end{equation}
where if a negative site was chosen one would replace the $\bm{n_0}, \mathfrak{p}$ in each numerator with $\bm{n_0},
\mathfrak{n}$. The equivalent expression for the cases without internal states has been presented in ref.
\cite{biroli2022number}.

Equation (\ref{eq: MNDVS_z}) represents the general formulation of the MNDVS for multi-state Markov walks, valid both
for unbounded or bounded domains. If the aforementioned condition is met ($ \widetilde{P}_{\bm{n},
\mathfrak{n}}^{(\mathfrak{n})}(\bm{n}, z) = \widetilde{P}_{\bm{0}, \mathfrak{p}}^{(\mathfrak{p})}(\bm{0}, z) $), i.e.,
the probability of a walker occupying its initial site is invariant to the location and site type of the initial
condition, we may simplify Eq. (\ref{eq: MNDVS_z}) further to 
\begin{align}
    \widetilde{\mathcal{M}}(z) &= \frac{1}{(1-z)\widetilde{P}_{\bm{0}, \mathfrak{p}}^{(\mathfrak{p})}(\bm{0}, z)}\sum_{\bm{n}}\left[\widetilde{P}_{\bm{n_0}, \mathfrak{p}}^{(\mathfrak{p})}(\bm{n}, z)+\widetilde{P}_{\bm{n_0}, \mathfrak{p}}^{(\mathfrak{n})}(\bm{n}, z)\right], \nonumber \\ 
                     &= \frac{1}{(1-z)^2\widetilde{P}_{\bm{0}, \mathfrak{p}}^{(\mathfrak{p})}(\bm{0}, z)},
    \label{eq: MNDVS_z1}
\end{align}
where we have used, in the second line, the property that the sum of the occupation probability over all the sites and states equals unity to
recover the well-known form of the MNDVS \cite{rws_on_latticesII,blumen1981energy}. For non-isotropic
lattices, e.g., in the presence of reflecting domains, one needs to employ Eq. (\ref{eq: MNDVS_z}) contrary to what is
erroneously implied in ref. \cite{LucaPRX}.

Upon substitution of the generating function of Eq. (\ref{eq: corr_t}) into Eq. (\ref{eq: MNDVS_z1}) and performing some
algebra, one arrives at 
\begin{equation}
    \mathcal{M}(t) = \frac{2\pi}{i}\! \oint_{|z|<1}\!\frac{1}{(1\!-\!z)^2}\left[\sum_{j=1}^{3}\left(\int_{-\pi}^{\pi}\int_{-\pi}^{\pi}\left[\mathbb{I}-z\bm{\lambda}(\xi_1, \xi_2)\right]^{-1}\cdot \bm{U}_{\mathfrak{p}}\text{d}\xi_1\text{d}\xi_2\right)_j\right]^{-1}\text{d}z.
    \label{eq: MSV_integral}
\end{equation}
While evaluating Eq. (\ref{eq: MSV_integral}) analytically is a formidable task, it may be calculated straightforwardly
via numerical procedures. To do so, the inner integral is computed via a two-dimensional cubature scheme
\cite{genz1980remarks}, the output of which is then fed into a numerical contour scheme, utilising standard fast-Fourier
techniques to compute the contour integral \cite{abate1992fourier} .
\begin{figure}[h]
    \includegraphics[width=\textwidth]{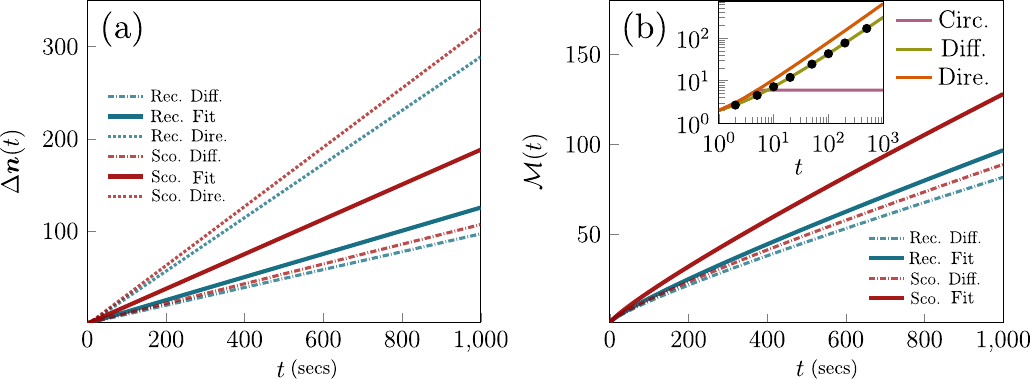}
    \caption{A comparison between the scout and recruit spreading statistics over the infinite honeycomb lattice with
    the mean squared displacement in (a) and the mean number of visited sites in (b). The estimated model parameters
    from data (i.e., $r$, $l$, $b$ and $c$) for both scouts and recruiters are given explicitly in Sec. 2. We display
    the curves with these parameters with the label `Fit' in the legend and compare them against the diffusive and
    directed limits. We take  the sojourn probability $c$ to be $c\approx 0.7594$ (scouts) and $c \approx 0.7821$
    (recruits), leading to $b=0$, and $r=l=(1-c)/2$ and $r=b=l = (1-c)/3$ for the directed and diffusive cases,
    respectively. In the inset of panel (b) we plot three limiting cases (with dimensionless time), each with $c=0$. The
    black dots show excellent agreement with the results obtained iteratively in ref. \cite{zumofen1982energy} for the
    diffusive walk, while the zero gradient past $t=6$ in the circular case is to be expected since the circular walker
    will repeat its journey around one hexagon for all time.}
    \label{fig: spreading}
\end{figure}

In Fig. \ref{fig: spreading} we plot the MSD and MNDVS for recruits and scouts, with comparison to diffusive and
directed spread. It is seen that the scouts explore the space more efficiently than the recruits and both behavioural
types explore quicker than a diffusive walker.   These results align with expectations, as animals with frontal and
bilateral sensory systems are known to exhibit more persistent directional movement compared to pure random walkers
\cite{Martin2013Behaviour, turchin1998quantitative}. Similarly, scouts are expected to be more effective explorers than
recruits, as the latter play a more critical role during food exploitation after scouts make initial patch discoveries.

\section{Search of $\mathcal{N}$ distinct walkers} We first present general theoretical findings in Sec. \ref{sec:
theor_ffpt} and \ref{sec:theory2_ffpt} pertaining to search with $\mathcal{N}$ independent random walkers, while in Sec.
\ref{sec: exp_ffpt} we consider model predictions pertaining to the ant experiment. Specifically, we consider the fFPT,
which is the probability that the first walker in the population reaches any target of a set $\mathcal{T}$ for the first
time at time $t$, in a heterogeneous population with individuals displaying different movement characteristics. The
exact theoretical development for the fFPT that we present below is particularly relevant now given the recent surge in
closed-form expressions for the first-passage and splitting probability generating function in a range of Markovian
\cite{LucaPRX,sarvaharmangiuggioli2023,giuggioli2024multi} and non-Markovian
\cite{marris2024persistent} walks. From the biological perspective, while in general ants share information in the food
gathering process \cite{AntEcology_chapter_foraging}, the application of the fFPT formalism is pertinent to our
experiments as recent work on \textit{A. senilis} has elucidated that during the exploration phase, i.e., before any
food is found, individual ants move largely independent of each other \cite{cristin2024spatiotemporal}, justifying the
assumption of independent walkers.

%While there is certainly information shared between ants in the food gathering process
%\cite{AntEcology_chapter_foraging}, recent work in the context of \textit{A. senilis} has elucidated that in the
%exploration phase (before any food is found) the ants in the arena move largely independent of the others
%\cite{cristin2024spatiotemporal}, justifying the assumption of independent searchers here.

\subsection{Theory of first first-passage processes in a heterogeneous population with multiple targets}\label{sec:
theor_ffpt} The probability that none of $\mathcal{N}$ independent random walkers have reached any site in the set of
targets $\mathcal{T}=\{\bm{s}_1, ..., \bm{s}_{\mathdutchcal{N}}\}$ is expressed via the survival probability of the
entire population $\mathcal{S}_{\bm{n}_0}(t; \mathcal{N})=\prod_{j=1}^{\mathcal{N}}\mathcal{S}^{(j)}_{\bm{n}_{0_j}}(t)$
(see e.g., refs. \cite{grebenkov2020single, lawley2020universal} and references therein), where
$\mathcal{S}^{(j)}_{\bm{n}_{0_j}}(t)$ is the survival probability of walker $j$ at time $t$ given it was located at
$\bm{n}_{0_j}$ at $t=0$. When the movement statistics of each walker are identical and the walkers start from the same
initial position, which represents the nest in our case, one has $\mathcal{S}_{\bm{n}_0}(t;
\mathcal{N})=\mathcal{S}_{\bm{n}_0}(t)^{\mathcal{N}}$. In continuous time, the direct relation of the survival
probability and the first-passage probability to any of the targets in the set $\mathcal{T}$, namely
$F_{\bm{n}_0}(\mathcal{T}, t) = - \frac{\text{d}}{\text{d}t} \mathcal{S}_{\bm{n}_0}(t)$, leads to the well-known
expression \cite{weiss1983order}
\begin{equation}
F_{\bm{n}_0}(\mathcal{T}, t;\{1,\mathcal{N}\}) = \mathcal{N}F_{\bm{n}_0}(\mathcal{T}, t)\mathcal{S}_{\bm{n}_0}(t)^{\mathcal{N}-1},
\label{eq: 1st_of_N_cont}
\end{equation} 
which represents the fFPT time of $\mathcal{N}$ independent and identical random walkers, all starting from $\bm{n}_0$.
Note that in continuous time Eq. (\ref{eq: 1st_of_N_cont}) implies that the probability of more than one walker reaching
the target for the first time at identical times is exactly zero. In discrete time, this is no longer the case since
events of simultaneous spatial coincidence at the target of more than one walker (for the first time) are possible. 

To allow for such a situation, we modify Eq. (\ref{eq: 1st_of_N_cont}) to 
\begin{equation}
    F_{\bm{n_0}}(\mathcal{T}, t;\{1,\mathcal{N}\}) = \sum_{k=1}^{\mathcal{N}}\binom{\mathcal{N}}{k}F_{\bm{n_0}}(\mathcal{T}, t)^{k}\mathcal{S}_{\bm{n_0}}(t)^{\mathcal{N}-k}, 
    \label{eq: 1st_of_N}
\end{equation}
where $\binom{n}{r}$ is the binomial coefficient. Equation (\ref{eq: 1st_of_N}) accounts for two distinct first-passage
events: (i) the target has been reached for the first time by exactly one walker and none of the others, and (ii) more
than one walker reaches the target for the first time. The event (i) is accounted for by the $k=1$ term of the
summation, which represents the case in Eq. (\ref{eq: 1st_of_N_cont}), while the contribution of the $k=2,3,...,
\mathcal{N}$ terms quantifies the probability of event (ii). For a given $k\geq 2$, there are $k$ walkers among the
$\mathcal{N}$ that may reach the target at the same time. As we do not know which member(s) of the population reached
the target, the binomial coefficient, multiplying each element of the series, takes into account all the possible ways
to pick $k$ walkers among the $\mathcal{N}$. The independence of the events then demands to take the product of the
probability for $k$ first-passage events with the probability that the remaining $\mathcal{N}-k$ walkers have not
reached the target. Note that while outcome (ii) encompasses a larger range of statistically realisable events, it is
generally much rarer than (i). However, under certain conditions event (ii) may become more frequent. For instance, with
a large number of persistent walkers the probability that more than one walker occupies lattice sites at equivalent
distance from the initial condition at the same time is non-negligible. 

In the case where the search involves more than one target ($\mathdutchcal{N}>1$) we extend the theory to account for
the splitting probability $F_{\bm{n}_0}(\bm{s}_i|\bm{s}_{\{1,…,\mathdutchcal{N}\}-\{i\}}, t)$, that is the probability
of a first first-passage event at target $\bm{s}_i$ conditioned on no member of the population reaching any other target
in the set $\bm{s}_{\{1,…,\mathdutchcal{N}\}-\{i\}}$ in the times up to and including $t$. In such a scenario one
decomposes the event (ii) into two mutually exclusive first-passage events: (iia) more than one walker simultaneously
reach target $\bm{s}_i$ and (iib) one  or more walkers reach target $\bm{s}_i$ while one or more walkers reach one of
the remaining sites in the target set $\bm{s}_{\{1,…,\mathdutchcal{N}\}-\{i\}}$. For mathematical convenience we take
the event (iia) as a success and (iib) as a failure, and we leave the derivation of a general formalism that accounts
also for events (iib) to a future publication. 

To derive the expression for the splitting probability we observe that by construction the survival probability is
dependent on a walker having not reached any target in the set. Since each successful walker must be at $\bm{s}_i$, a
replacement of the first-passage in Eq. (\ref{eq: 1st_of_N}) with the single walker splitting probability to $\bm{s}_i$
provides the required extension:
\begin{align}
    F_{\bm{n_0}}(\bm{s}_i&|\bm{s}_{\{1,…,\mathdutchcal{N}\}-\{i\}}, t;\{1,\mathcal{N}\})   \!=\! \sum_{k=1}^{\mathcal{N}}\binom{\mathcal{N}}{k}F_{\bm{n_0}}(\bm{s}_i|\bm{s}_{\{1,…,\mathdutchcal{N}\}-\{i\}}, t)^{k}\mathcal{S}_{\bm{n_0}}(t)^{\mathcal{N}-k}.
     \label{eq: splitting_1st_of_N}
 \end{align}
\sloppy Equation (\ref{eq: splitting_1st_of_N}) represents the time-dependent first-splitting probability for one or
more walkers among $\mathcal{N}$, starting at $\bm{n}_0$, to reach target $\bm{s}_i$ before any other in the set of
$\mathdutchcal{N}$ targets. As the event (iib) is excluded from all possible outcomes making up
$F_{\bm{n}_0}(\bm{s}_i|\bm{s}_{\{1,…,\mathdutchcal{N}\}-\{i\}}, t;\{1,\mathcal{N}\})$, one needs to keep in mind that
for $\mathcal{N}\neq 1$, $\sum_{i=1}^{\mathdutchcal{N}} F_{\bm{n}_0}(\bm{s}_i|\bm{s}_{\{1,…,\mathdutchcal{N}\}-\{i\}},
t;\{1,\mathcal{N}\})\neq F_{\bm{n}_0}(\mathcal{T}, t;\{1,\mathcal{N}\})$.

We now further extend the notion of search for a heterogeneous population to allow for the different movement patterns
and/or different initial conditions of the individuals. We do so by partitioning the population into $h \leq
\mathcal{N}$ subsets, where each subset is comprised of $\mathcal{N}_h$ walkers with the same movement statistics such
that $\sum_{h}\mathcal{N}_h = \mathcal{N}$ and $\mathcal{N}_1 \geq \mathcal{N}_{2} \geq ... \geq \mathcal{N}_h$. By
considering all permutations of possible successful search outcomes, the generalisation of Eq. (\ref{eq: 1st_of_N}), to
a heterogeneous population, denoted by the superscript $\mathfrak{h}$, is given by  
\begin{align}
F_{\bm{n}_0}^{(\mathfrak{h})}(\mathcal{T}, t;\{1,\mathcal{N}\}) &=  \sum_{i=1}^{h}\sum_{k_i=1}^{\mathcal{N}_i}\binom{\mathcal{N}_i}{k_i}F_{\bm{n}_{0_i}}^{(i)}(\mathcal{T}, t)^{k_i}\prod_{j=1}^{h}\mathcal{S}_{\bm{n}_{0_j}}^{(j)}(t)^{\mathcal{N}_j-\delta_{i, j}k_j} \nonumber \\ 
&+\sum_{k_1=1}^{\mathcal{N}_1} 
\cdots \sum_{k_h=1}^{\mathcal{N}_h}\prod_{i=1}^{h}\binom{\mathcal{N}_i}{k_i}F_{\bm{n}_{0_i}}^{(i)}(\mathcal{T}, t)^{k_i}\mathcal{S}_{\bm{n}_{0_i}}^{(i)}(t)^{\mathcal{N}_i-k_i } ,
\label{eq: nonI1st_of_N}
\end{align}
which reduces to Eq. (\ref{eq: 1st_of_N}) when $\mathcal{N}_1 = \mathcal{N}$ ($\mathcal{N}_{i} = 0$, $\forall i > 1$) as
we define the nested sum in the second term to be zero in this case. This makes sense physically as the first term
represents the case of one or more walkers from the same subset reaching the target at the same time (summed over each
subset), while the second term represents the cases where a mixed population of walkers reaches the target for the first
time. The splitting probability expression with a heterogeneous population may be derived via an analogous procedure to
the homogeneous case, i.e., by replacing the first-passage probabilities in Eq. (\ref{eq: nonI1st_of_N}) with the
splitting probabilities for the type of walkers belonging to the subset population.

The general expression in Eq. (\ref{eq: nonI1st_of_N}) is important both from a theoretical and an applied perspective.
The theoretical advance lies in the ability to study first-passage processes to an arbitrary number of targets for
multiple independent walkers, with each individual following their own movement statistics. The applied convenience
stems from the computational cost in evaluating $F_{\bm{n}_0}^{(\mathfrak{h})}(\mathcal{T}, t;\{1,\mathcal{N}\})$:
despite its appearance Eq. (\ref{eq: nonI1st_of_N}) has time complexity on the same order as the one walker case since
one determines the survival probability for each subset $h$ prior to evaluating the full summations.

\subsubsection{First first-passage probability in bounded domains}
\label{sec:theory2_ffpt}
Most studies of the fFPT with $\mathcal{N}$ walkers have focused on large homogeneous populations of walkers searching
in unbounded space \cite{weiss1983order,lawley2020universal, lawley2020probabilistic, meerson2015mortality}. Here
instead, guided by the observations of \textit{A. senilis}, we consider a small population of $\mathcal{N}$
($\mathcal{N}\leq 10$) individuals with unequal movement statistics. As an illustrative example we consider a search
process in bounded periodic space and study two cases: a population comprised of both strongly persistent and diffusive
walkers on the square lattice, and a population of correlated walkers, representing scouts and recruits, on a honeycomb
lattice. 

For the square lattice case we take the well-known diffusive periodic propagator (see e.g., refs. \cite{LucaPRX,
barry_hughes_book, montroll1956random}) and the corresponding correlated propagator given by Eq. (14) of ref.
\cite{marris2024persistent}. For correlated motion on the bounded honeycomb lattice we introduce a periodic boundary on
Eq. (\ref{eq: corr_t}) via standard procedures (see e.g., \cite{barry_hughes_book}) to find
\begin{equation}
\widetilde{\bm{P}}_{\bm{n}_0, m_0}^{(p)}(\bm{n}, z) = \frac{1}{N^2}\sum_{k_1=0 }^{N-1}\sum_{k_2=0 }^{N-1}\text{e}^{-\frac{2\pi i}{N} \bm{k}\cdot[\bm{n}-\bm{n}_0]^{\intercal} }\left[\mathbb{I}-z\bm{\lambda}\left(\frac{2\pi k_1}{N},\frac{2\pi k_2}{N}\right)\right]^{-1}\cdot U_{m_0},
     \label{eq: per_corr_honey}
\end{equation}
where the inversion to the time-domain is trivial and the superscript $(p)$ denotes that the domain is bounded
periodically. 

From ref. \cite{marris2024persistent}, the generating function of the single walker first-passage distribution for the
correlated RW is found via 
\begin{equation}
 \widetilde{F}_{\bm{n}_0}(\mathcal{T}, z) = \sum_{j=1}^{M}\sum_{i = 1}^{S}\alpha_{m_{\bm{s}_i}}\frac{\det[\mathbb{H}^{(i)}(\bm{n}_0, m_{0_j}, z)]}{\det[\mathbb{H}(z)]},
 \label{eq: multi_tar_FP}
\end{equation}
 and the splitting probability is given by 
\begin{equation}
 \widetilde{F}_{\bm{n}_0}(\bm{s}_{\{i\}}|\bm{s}_{\{1,…,\mathdutchcal{N}\}-\{i\}}, z) = \sum_{j=1}^{M}\alpha_{m_{\bm{s}_i}}\frac{\det[\mathbb{H}^{(i)}(\bm{n}_0, m_{0_j}, z)]}{\det[\mathbb{H}(z)]},
 \label{eq: multi_tar_split}
\end{equation}
\sloppy where $\mathbb{H}(z)_{l,k} = \alpha_{m_{\bm{s}_k}}\widetilde{P}_{\bm{s}_k, m_{\bm{s}_k}}(\bm{s}_l, m_{\bm{s}_l},
z)$, $\mathbb{H}(z)_{k,k} = \alpha_{m_{\bm{s}_k}}\widetilde{P}_{\bm{s}_k, m_{\bm{s}_k}}(\bm{s}_k, m_{\bm{s}_k}, z)$ and
$\mathbb{H}^{(i)}(\bm{n}_0, m_0, z)$ is the same, but with the $i^{\text{th}}$ column replaced with $
\mathdutchcal{F}(n_0, m_{0_j}, z) = \alpha_{m_{0_j}}\left[\widetilde{P}_{\bm{n}_0, m_{0_j}}(\bm{s}_1, m_{\bm{s}_1}, z),
..., \widetilde{P}_{\bm{n}_0, m_{0_j}}(\bm{s}_\mathdutchcal{N}, m_{\bm{s}_{\mathdutchcal{N}}}, z) \right]^{\intercal}$. 
%The splitting probability for the Markovian diffusive case may be found in Eq. (11) of ref. \cite{giuggioli2022spatio}.
The notation $\alpha_{m_{0_j}}P_{\bm{n}_0, m_{0_j}}(\bm{n}, m, t) = \alpha_{m_{0_j}}\bm{e}_m^{\intercal}\cdot
\bm{P}_{\bm{n}_0}(\bm{n}, t)\cdot \bm{e}_{m_{0_j}}$, is the occupation probability that the walker occupies site
$\bm{n}$ and state $m$ given the initial condition $(\bm{n}_0, m_{0_j})$, which is weighted by $\alpha_{m_{0_j}}$.
Presently, due to the assumption of each direction being equi-probable at time zero $\alpha_{i} = 1/3$, $\forall i$.
Moreover, since each site can be accessed via three distinct internal states, to represent a fully absorbing trap we set
the probability of leaving each internal state in the trap site to zero. For example, to place a target in a positive
site, absorbing traps in the states $1, 2$ and $3$ are required. Hence, for all cases presented here, and elsewhere in
this chapter, the  target set $\mathcal{T}$ is comprised of  sets of three target states for each physical site.
Henceforth, for ease of notation, when discussing specific cases we denote the set of targets with only the physical
location(s) at site(s) $\bm{s}_i$. The splitting probability for the Markovian case for the diffusive subset can be found in Eq. (11) of ref. \cite{giuggioli2022spatio}.

\begin{figure}[h!]
    \centering
    \includegraphics[width=\linewidth]{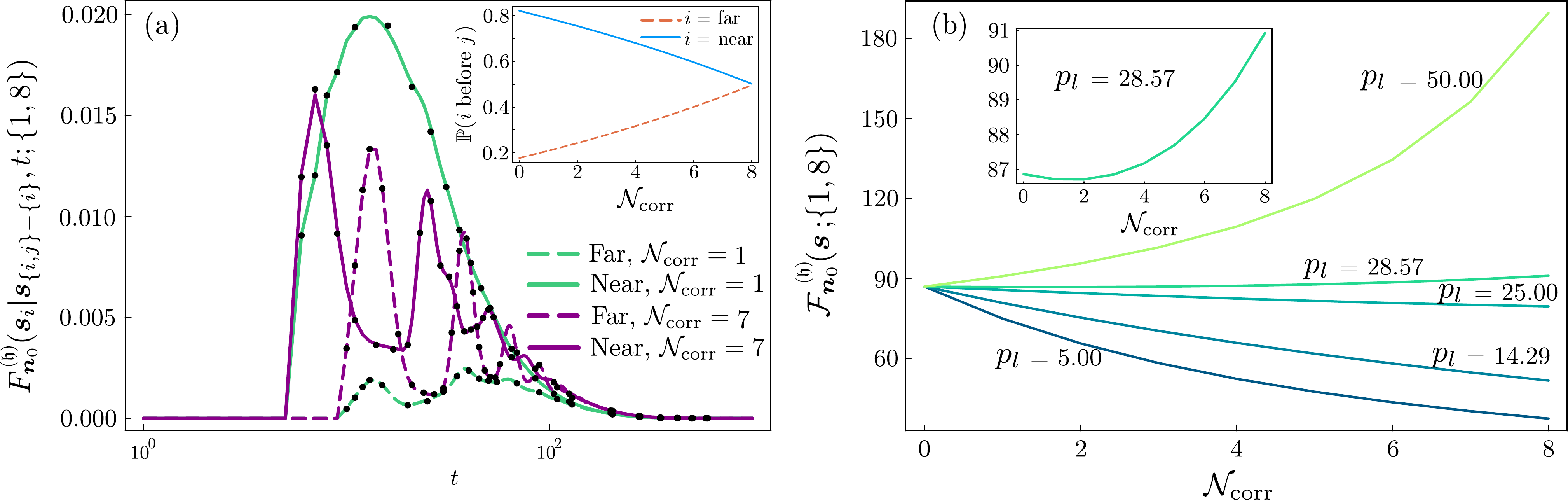}
    \caption{The fFPT for different heterogeneous populations of random walkers on the periodic square lattice. The
    population is comprised of $\mathcal{N} = \mathcal{N}_{\text{corr}}+\mathcal{N}_{\text{diff}}=8$ random walkers. In
    panel (a) we evaluate Eq. (\ref{eq: nonI1st_of_N}) with the first-passage terms replaced with the splitting
    probability for the single persistent walker given by Eq. (\ref{eq: multi_tar_split}). The population starts at
    $\bm{n}_0 = (10,10)$ and is searching for a nearby target at $\bm{s} = (7,7)$ and a far away target at $\bm{s} =
    (1,9)$ in a domain size $N_1 \times N_2 = 20\times 20$. Hence, with the square lattice requiring four internal
    states to represent the possible previous movement directions, and the system contains two physical target sites, we
    have $\mathdutchcal{N} = 8$ target states. For the diffusive population we take $q = 0.95$, while the persistent
    population has a persistence length $p_l = 20$. The inset shows the overall probability that each target is found
    before the other (the summation of the splitting probability over time) as a function of $\mathcal{N}_{\text{corr}}$
    for the same parameter regime. Panel (b) shows the MfFPT, $\mathcal{F}_{\bm{n}_0}^{(\mathfrak{h})}(\mathcal{T}; \{1,
    \mathcal{N}\} = \sum_t\,t F_{\bm{n}_0}^{(\mathfrak{h})}(\mathcal{T}, t; \{1, \mathcal{N}\})$,  as a function of
    $\mathcal{N}_{\text{corr}}$ with varying persistent lengths all chosen past the optimum for this system $p_l \approx
    4.76$. Note that the diffusive persistence length is $p_l = 4(3q)^{-1}$ and for the diffusive members of the
    population we take $q=0.95$ ($p_l \approx 1.40)$. In this case we choose a domain size of $N_1 \times N_2 = 14\times
    14$, $\bm{n}_0 = (7,7)$ and $\bm{s} = (14,14)$ meaning $\mathdutchcal{N} = 4$. Dots represent the result of $10^{6}$
    stochastic simulations.}
    \label{fig: thoer_ffpt}
\end{figure}

We plot in Fig. \ref{fig: thoer_ffpt} the fFPT for different heterogeneous populations on the square lattice. To present
results on the square lattice it is convenient to introduce the so-called persistence length
\cite{tejedor2012optimizing}, that is the average number of steps taken in the same direction between turns. On the
square lattice it is defined as $p_l = [b+2\tau]^{-1}$, where $b$ is the probability of backtracking and $2\tau$ the
probability of turning left or right (probability $\tau$ in each direction). In Fig. \ref{fig: thoer_ffpt}(a) we
consider a heterogeneous population of eight walkers searching on a domain of $20\times 20$ lattice sites. The domain
contains two targets, one close-by to the initial position and the other far away and just off the persistence path,
(see  caption of Fig. \ref{fig: thoer_ffpt} for the exact locations). By increasing the proportion of persistent walkers
in the population (inset Fig. \ref{fig: thoer_ffpt}(a)) the overall probability $\mathbb{P}$ of finding the far target
increases until it is almost equi-probable as many trajectories head in a straight-line out of the initial condition
bypassing the closer target. 

In panel (a) one notice clear qualitative differences between the splitting probabilities of the search with different
populations. Persistence may lead to oscillatory behaviour in the first-passage distribution of one walker
\cite{marris2024persistent}, something that is greatly suppressed as the movement parameters approach the diffusive
regime. These oscillations, which are not present with diffusive walkers, have low amplitude when one persistent walker
is introduced, while they increase in regularity and amplitude as the number of persistent walkers increase.

To gain further insights on the role of persistence in the first-passage processes we interpret our findings also in
terms of the mean fFPT (MfFPT), for which it is known that there is an optimal persistence length $p_l$, whose value
depends on the search parameters, that minimises the global search time on the square lattice
\cite{tejedor2012optimizing}. If the persistence parameter greatly surpass $p_l$, one may find very high mean
first-passage time if the target is off the persistence path. Here in Fig. \ref{fig: thoer_ffpt}(b) we show persistence
lengths starting from just past the optimum for the chosen system ($p_l \approx 4.76$). In such a parameter regime, an
increase in the persistence length hinders the search and weakens the effect of adding more persistent walkers until a
population of diffusive walkers outperforms a population of correlated walkers. Around the level of persistence which
tips the walker beyond the diffusive efficacy (shown presently as $p_l = 28.57$), the MfFPT becomes non-monotonic as a
function of the number of correlated walkers in the population as shown in the inset of panel (b).

\begin{figure}[h!]
    \centering
    \includegraphics[width=\linewidth]{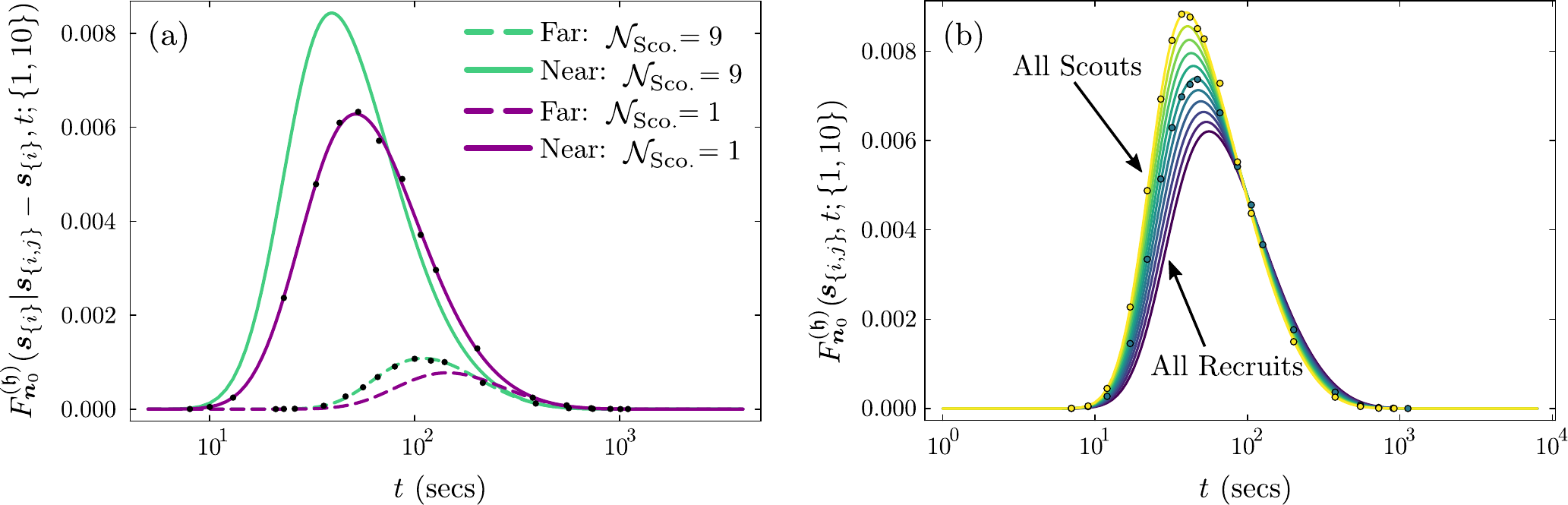}
    \caption{The splitting probability to either site (a) and first first-passage probability to both sites (b) of an
    $\mathcal{N} = \mathcal{N}_{\text{Sco.}} +\mathcal{N}_{\text{Rec.}} = 10$ heterogeneous population of scouts and
    recruits (see sec. 2 for parameter values) in an $N \times N = 12 \times 12$ (288 sites) periodic honeycomb lattice.
    For all curves the walkers start in the positive site of $\boldsymbol{n}_0 = (6, 6)$. The target set $\mathcal{T}$
    is comprised of the three internal states in a nearby target in the negative site of $\boldsymbol{s}_1 = (9,9)$ and
    the three internal states in the far target in the positive site of $\boldsymbol{s}_2 = (11, 12)$ (see caption of
    Fig. \ref{fig: lattice_schematic} for definition of positive and negative sites). In both panels lines represent
    theoretical results (Eq. (\ref{eq: nonI1st_of_N}) evaluated via Eqs. (\ref{eq: multi_tar_FP}) and (\ref{eq:
    multi_tar_split}), respectively) while dots are the result of $10^6$ stochastic simulations, which we omit from some
    of the curves for visual clarity.}
    \label{fig: honey_theor_ffpt}
\end{figure}

For the experimentally relevant case we consider scouts and recruits in a honeycomb lattice and we plot the
first-passage dynamics in Fig. \ref{fig: honey_theor_ffpt}. In panel (a) we show the splitting probability to a nearby
and far target. In such a case, the closest target is always more likely to be found regardless of the proportion of
scouts to recruits, which is not only expected to be more energetically efficient, but it also matches the exploratory
patterns observed in different ant species \cite{pol2024liquid, martin2024}. Panel (b) shows the entire fFPT
distribution for each ratio of scouts to recruits. Since an increase in the proportion of scouts allows the space to be
explored more efficiently the so-called direct trajectories become more effective causing the mode of the fFPT
distribution to shift upwards and towards the left as it becomes more probable to find the target at shorter times. 

\subsection{Experimentally relevant search on the Y-maze arena}\label{sec: exp_ffpt}

We now consider a model which faithfully represents the domain size and boundaries; food and nest locations; and the
number of ants in the system for each experimental realisation. Since the ant population in the arena was dynamic, with
ants constantly entering and leaving the nest, we estimated the number of ants by considering how many ants were
exploring the arena at the time when any food target was found (end of exploration phase) as the static number
$\mathcal{N}$ for the model. The experimental set-up is modelled as a honeycomb lattice confined in a rectangular space,
with hard, `bouncing' reflecting walls, which causes persistent walkers to change direction at the boundary and continue
its motion in the opposite direction \cite{marris2024persistent}. 

To implement a reflecting boundary condition on the honeycomb lattice, note that although there are positive and
negative states occupying the unit cells that lie at the edge of the domain, if the site is not at the corner, only one
of the two states will interact with the boundary creating the so-called zig-zag boundary condition
\cite{marris2023exact,henry2003random}. To elucidate, consider Fig. \ref{fig: lattice_schematic} (iii) where, although
the positive site labelled y, lies in unit cell at the edge of the square lattice, all three movement options are
available to the walker leaving the site. In contrast, travelling horizontally East from the negative site labelled e
would cause a reflection to occur. In general, for the $\bm{n} = (N_1, n_2)$ boundary, the walker may only hit the
boundary if it leaves a negative site, while at the $\bm{n} = (n_1, N_2)$ boundary it is the positive site that
interacts with the boundary, meaning at $\bm{n} = (N_1, N_2)$, both sites are part of the reflecting boundary. 

To implement this zig-zag reflecting boundary condition via a fully analytical procedure is a formidable task when the
movement is not diffusive. To proceed we use a semi-analytical iterative procedure
\cite{marris2024persistent,gueneau2024siegmund}, which has been used to study persistent motion with complicated
boundary conditions where the analytical representation of the occupation probability is not available. This procedure
requires the decomposition of the movement dynamics into matrices that govern only the jumps from each state $i$ to
every other state $j$, denoted by $\mathbb{M}(i, j)$. In our case this leads to 36 sparse transition matrices, of size
$N_1N_2 \times N_1N_2$. More concretely, the matrices that govern movement in the bulk have $1/N_1N_2$ ($\approx0.31\%$)
non-zero elements, while those governing movement on the boundary have $1/N_1^2N_2$ ($\approx 0.022\%$) or $1/N_1N_2^2$
($\approx 0.0135\%$) non-zero elements. Hence, all matrices benefit from the use of sparse methods leading to efficient
computation. See Appendix 3 for a detailed discussion of the transition matrices.

Encoded into $\mathbb{M}(i,j)$ is also the probability of absorption, i.e., the outgoing probabilities of the trap site,
i.e., the site in which the first food source is found, sum up to $1-\rho$, where $\rho$ is the probability of
absorption (finding the food) at the target. Experimentally, we estimated $\rho$ by measuring the result of each
ant-food interaction. Namely, given an interaction event (i.e., an ant finds a food item), we consider a success
(absorption) the ant collecting the food item, or a failure otherwise. Subsequent interactions were considered
independent when the ant had walked away from the food for two or more nodes. The sum of successes divided by the number
of interactions results in an empirical estimate of $\rho \approx 0.82$.

Once the elements of transition matrices are generated, the Master equation for the bounded process can be written as a
set of six coupled equations as
\begin{equation}
    \begin{aligned}
            \bm{P}(j, t+1)  &= \sum_{i=1}^6\mathbb{M}(i,j)\cdot\bm{P}(i, t), \; \; (\forall j \in \{1, 6\}),
        \end{aligned}  
        \label{eq: mat_ME}
\end{equation}
where $\mathbb{M}(i,j)$ is the probability of jumping from state $i$ to state $j$. Then via
$\mathcal{S}_{\bm{n}_0}(t)=\sum_{i=1}^{6}\sum_{n =1}^{N_1N_2}\bm{P}(i,t)_n$, and $F_{\bm{n}_0}(\mathcal{T}, t) =
\mathcal{S}_{\bm{n}_0}(t-1) - \mathcal{S}_{\bm{n}_0}(t) $ , where the inclusion of $\rho \neq 0$ ensures $0 \leq
\mathcal{S}_{\bm{n}_0}(t) \leq 1$, one has all the information required to evaluate the fFPT via Eq. (\ref{eq:
nonI1st_of_N}). Once the entire fFPT distribution is known, the MfFPT and mode fFPT time are extracted trivially via
$\mathcal{F}_{\bm{n}_0}(\mathcal{T}; \{1, \mathcal{N}\}) = \sum_{t}tF_{\bm{n_0}}(\mathcal{T}, t;\{1,\mathcal{N}\})$ and
$\mathfrak{F}_{\bm{n}_0}(\mathcal{T}; \{1, \mathcal{N}\}) = \argmax\left[F_{\bm{n_0}}(\mathcal{T},
t;\{1,\mathcal{N}\})\right]$, respectively. 

\section{Results}\label{sec: results} 
\begin{figure}[th]
    \centering    \includegraphics[width=0.8\linewidth]{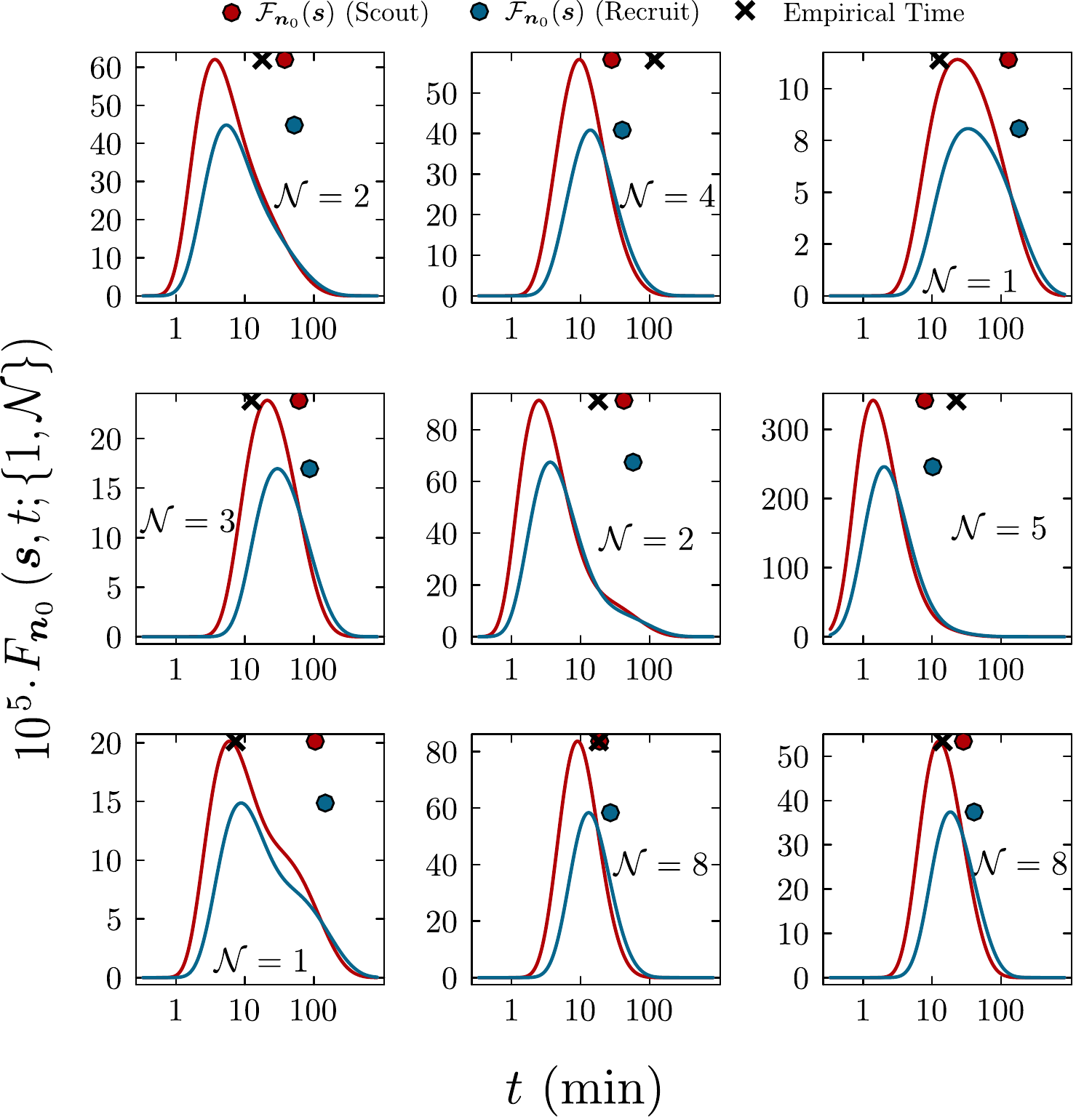} \caption{The first first-passage time
    distributions in Eq. (\ref{eq: 1st_of_N}),  evaluated with the output of Eq. (\ref{eq: mat_ME}), for a population of
    scouts (red) and recruits (blue) modelling experiments one to nine in the honeycomb arena with hard reflecting
    walls. Physical target locations, $\bm{s}$, are given in Appendix 4, and the set of targets $\mathcal{T}$ is
    comprised of the three internal states within. Results for each experiment are shown in order (from one to nine)
    starting from the top left corner panel and ending at the bottom right corner panel, moving from one panel to the
    next one by rows. Crosses denote the empirical times for the food item found while red (scouts) and blue (recruits)
    dots indicate the mean value of the fFPT distributions. The $x$-axis has been scaled by a division of 60 to give
    results in minutes.}
    \label{fig: exp_v_theor}
\end{figure}

We have analysed the full fFPT distribution for homogenous populations of either
scouts or recruits, using the same food placements (see Appendix 4 for specific sites) and the number of agents as in
the empirical data (see Fig.~\ref{fig: exp_v_theor}). For all distributions, the mean and modal times were lower for the
scouts than the recruits. With the exception of experiments two and six, the empirical times of first food item capture
(black crosses) falls in a timescale, which may be characterised by the mode of the theoretical distribution, showing
that generally, the empirical time of food discovery corresponds to the direct regime of the theoretical distributions.
This region, comprised of the time scale from the first deviation from zero up to and just past the mode
\cite{godec2016first}, is characterised by the set of trajectories that deviate minimally from paths that connect
directly the initial condition to the target. Figure \ref{fig: exp_v_theor} also shows that in general, including more
walkers to the population leads to a narrower fFPT distribution with sharper tails and shapes that are more symmetric
around the mode. In other words, the faster the tails decay, the smaller the difference between the mean and the mode.

\begin{figure}
    \centering    \includegraphics[width=\linewidth]{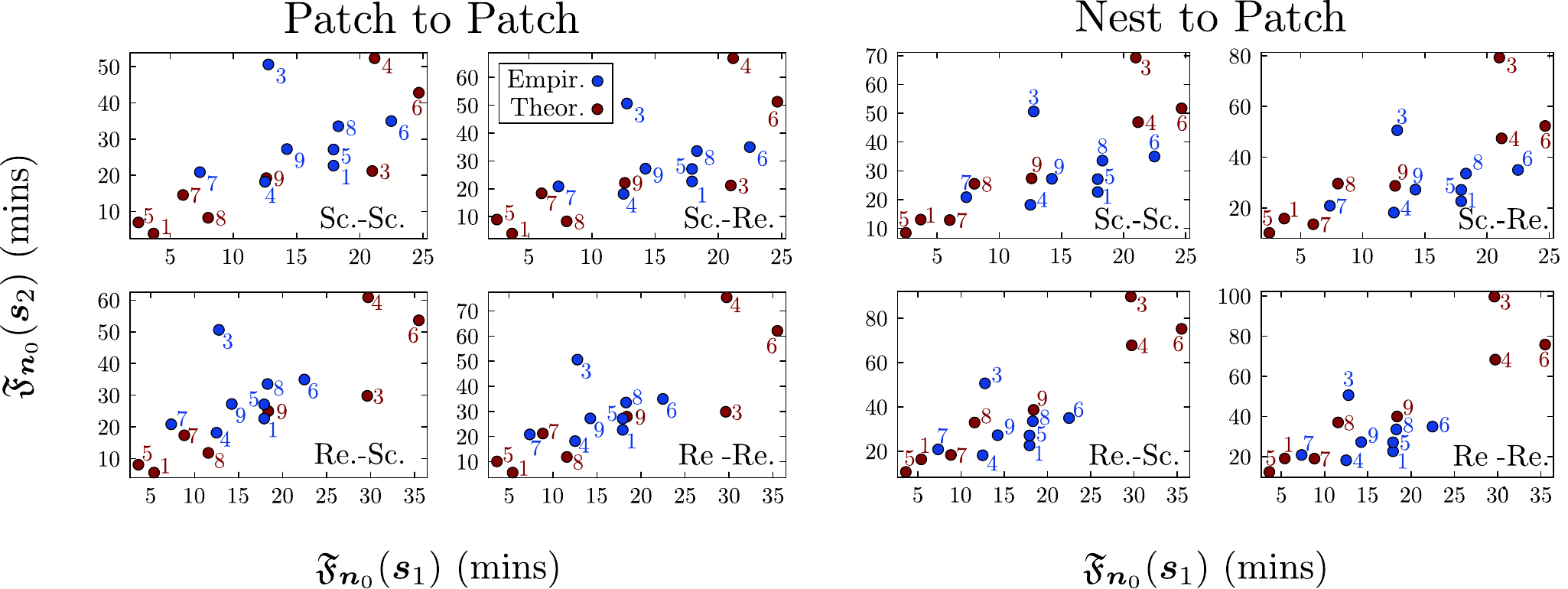} \caption{Relationship between the fFPT of the
    first patch (x-axis) and the second patch (y-axis). We considered two different foraging strategies: using
    previously found resources as starting points to search for new food targets (Patch to Patch, left panels) and
    strict central place foraging, wherein ants go back to the nest before searching for subsequent food targets  (Nest
    to Patch, right panels). We allow the entire population to switch between scout (Sc.) and recruit (Re.) upon
    reaching the first target in the patch to patch case or when returning from the nest in the nest to patch scenario.
    As the theoretical values we show the mode of the fFPT, $\mathfrak{F}_{\bm{n}_0}(\bm{s}; \{1, \mathcal{N}\})$ in
    red, and the first food capture time in blue. The numbers relate to the experimental setup, i.e., target placements
    and number of ants, given explicitly in Appendix 4 and Fig. (\ref{fig: exp_v_theor}), respectively. Due to the large
    emprical value in experiment two, we do not plot this case here.}
    \label{fig: fp1fp2}
\end{figure}

With ant colonies facing scenarios with multiple food sources in nature, we have explored how switching behaviour, e.g.,
from scout to recruit, may affect search efficiency in a landscape with multiple targets (Fig.~\ref{fig: fp1fp2}). For
each experiment we assume a homogeneous population comprised of $\mathcal{N}$ ants observed experimentally in
the laboratory set-up (written explicitly in Fig.~\ref{fig: exp_v_theor}), starting either as scouts or recruits. By
assuming that the entire population may modify searching patterns upon finding the first target or when returning back
to the nest, we address the question of behavioural switching in two possible foraging strategies; (i) nest-to-patch
foraging, i.e., strict central place foraging, in which ants go back to the nest before searching for the next target,
and (ii) patch-to-patch foraging, where the ants use the previously discovered patch as a starting point to search for
new food targets. The theoretical time for scenario (i) is defined as $\mathfrak{F}_{\bm{n}_0}(\bm{s_2}; \{1,
\mathcal{N}\}) = \mathfrak{F}_{\bm{n}_0}(\bm{s_1}; \{1, \mathcal{N}\})+\mathfrak{F}_{\bm{s_1}}(\bm{n}_0; \{1,
\mathcal{N}\})+\mathfrak{F}_{\bm{n}_0}(\bm{s_2}; \{1, \mathcal{N}\})$, with $\bm{s}_1$ being the first found patch, and
$\bm{s}_2$ the second one, while we define the theoretical time for scenario (ii) as $\mathfrak{F}_{\bm{n}_0}(\bm{s_2};
\{1, \mathcal{N}\}) = \mathfrak{F}_{\bm{n}_0}(\bm{s_1}; \{1, \mathcal{N}\})+\mathfrak{F}_{\bm{s_1}}(\bm{s_2}; \{1,
\mathcal{N}\})$. Note that since the process is bounded by reflecting walls, the first first-passage probability is not
symmetric with respect to the initial condition and target i.e., $F_{\bm{n}_0}(\bm{s}, t; \{1, \mathcal{N}\})\neq
F_{\bm{s}}(\bm{n}_0, t; \{1, \mathcal{N}\})$. 

To interpret all experiments together, we show the overall trend in Fig.~\ref{fig: fp1fp2} where we see that scouts
searching for the first patch reduced the time required for finding the targets, i.e., smaller modal fFPT, regardless of
the foraging strategy (top panels), while switching from scout to recruit (Sc.-Re.) did not improve search efficiency in
any of the setups. In agreement with previous results (Figs. \ref{fig: thoer_ffpt} and \ref{fig: exp_v_theor}), recruits
starting the search resulted in larger modal fFPT irrespective of the context or the behavioural switching
(Fig.~\ref{fig: fp1fp2}, bottom panels). This was particularly evident in some spatial configurations, i.e., setups 3, 4
and 6, wherein the fFPT departed significantly from the empirical distribution.

\begin{figure}[!ht]
    \centering
    \includegraphics[width = \textwidth]{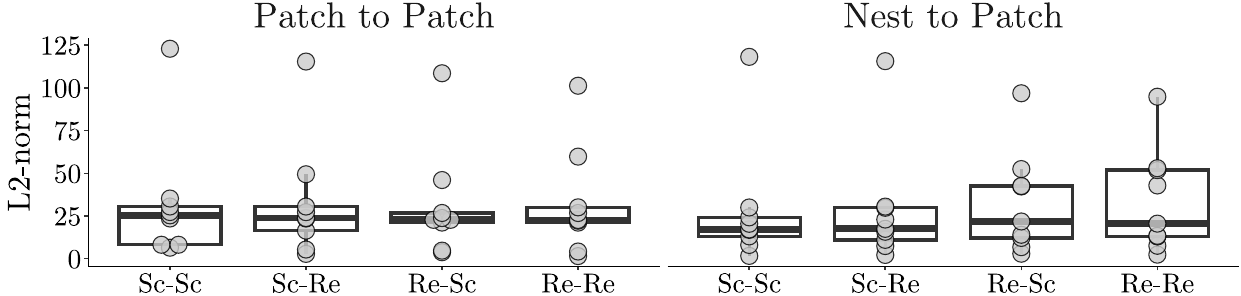}
    \caption{We illustrate the difference between empirical and theoretical data for each combination of foraging
    strategy and behavioural switch (Fig.~\ref{fig: fp1fp2}). We measured the Euclidean distance (L2 norm, y-axis)
    between the mode of the theoretical distribution and the corresponding empirical data point. The results are shown
    as box-and-whisker plots. The box delimits the interquartile range (IQR), compressed between the first (q1) and the
    third quartile (q3), with the 50\% percentile depicted as a thick line inside the box. The upper whisker is
    calculated as $\text{min}(\text{max}(\text{data}), \text{q3} + 1.5 \cdot \text{IQR})$, and the lower whisker is
    calculated as $\text{max}(\text{min}(\text{data}), \text{q1} - 1.5 \cdot \text{IQR})$. }
    \label{fig: l2_norm}
\end{figure}

We calculated the Euclidean distance between the mode of the theoretical distribution and the corresponding first
capture time in each empirical setup. The results shown in Fig.~\ref{fig: l2_norm} indicate that \textit{A. senilis}
likely uses a central-place foraging strategy (i.e., Nest to Patch), with scouts initially searching for the first
patch. The results are similar regardless of whether a behavioural switch from scout to recruit (Sc.-Re.) occurs after a
return to the nest following the discovery of the first patch or whether scouts continue the search to the second patch (Sc.-Sc.).
Nonetheless, the trend observed in Fig.~\ref{fig: l2_norm} is not statistically supported by pairwise Wilcoxon tests,
and the figure highlights a heterogeneous distribution of data points, that is certain setups reveal close alignment between
theoretical and experimental behaviours, while others do not. This is consistent with the analyses carried out in Figs.
\ref{fig: exp_v_theor} and \ref{fig: fp1fp2}.

\section{Discussion}\label{sec: discussion}
%Discuss the fact that we don't know whether recruits may also play some exploratory role which justifies exploring both
%types of movement modes in relation to exploratory behaviour.

% Escaping the average/mean as the 'ground truth', taking other measures such as modes or medians, which are more robust
% and arguably more suitable to quantify and understand animal behaviour (highly variable).
When collectively foraging, scouts and recruits are often identified in different ant species. They are commonly
distinguished by their positions relative to the nest, but recent studies of the ant \textit{A. senilis} have shown that
scouts and recruits are not only located in different positions relative to the nest but also move \cite{pol2024liquid}
and behave \cite{martin2024} differently. The dynamics of these two types of ants in group or mass recruitment
strategies are often complex and it is not easy to identify whether these are fixed or switching roles. In this chapter,
we studied the macroscopic consequences of locally different movement of both scouts and recruits in terms of their
exploratory and search efficiency. In particular, we studied the mean number of sites visited and mean squared
displacement for both ant roles, and also explored combinations of both types of ants in central place and non-central
place foraging scenarios. The latter allowed us to assess differences in foraging efficiency and identify what are the
more realistic scenarios when compared to empirical observations.

As the theoretical predictions using the nest to patch model aligns more closely with the empirical times than the patch to patch model,  (Fig.~\ref{fig: fp1fp2} and \ref{fig: l2_norm})
there are indications that a central place foraging strategy is
undertaken by the ants. These results suggest a temporal order in foraging, wherein ants first need to find a patch,
convey the information to the rest of the colony on their way back to the nest, e.g. by depositing scent, to start patch exploitation, and then
restart the search. In this manner the solution of the patch exploration-exploitation tradeoff is a sequential process in time,
where the exploration effort and the patch distance to the nest are the main drivers to forage efficiency. However, when
there are multiple food patches, ant colonies may parallelise the exploitation as a function of the exploration effort
of the colony, i.e., the number of scouts, and the initial positions from where scouting behaviour starts
\cite{pol2024liquid}. 

Moreover, Fig.~\ref{fig: fp1fp2} shows that scouts switching to recruit behaviour did not improve the colony performance
when searching for multiple patches. In fact, in all  exploration (Fig. \ref{fig: spreading}) and search (Figs.
\ref{fig: honey_theor_ffpt} and  \ref{fig: exp_v_theor}) scenarios studied, the scouts always outperformed the recruits.
This finding is supported by past theoretical investigations on square lattices showing that introducing persistence
into the motion of a searcher can lower searching times in comparison to a diffusing motion
\cite{tejedor2012optimizing}. Since our results show that recruits are statistically close to being diffusive while
scouts show persistent motion, one expects the scouts to be more efficient. Possible future work in this context
includes using the results on the honeycomb persistent random walk presented here to determine the optimum level of
persistence on the honeycomb lattice to assess the overall efficiency of the scouts' search.

Since all our results point to recruits being less efficient explorers than scouts, we question whether recruits have
any role in exploration. If diffusive or sub-diffusive searching by recruits around an area is important in detecting
nearby food around already-discovered patches,  recruits may have less of a role in explorative tasks. Because of their
motion, recruits often end up aggregated, maintaining a minimum group cohesion in certain areas, which might be crucial
for rapid information transfer and exploitation of known resources. To provide conclusive answers to these questions,
further research is needed to explore the roles that different movement patterns and switching behaviours could play in
exploration-exploitation tradeoffs while foraging.

%raising an open question about their specific role in foraging.
From a theoretical standpoint there has been a wealth of studies on the scaling of the MfFPT when the number of walkers
is large and homogeneous \cite{weiss1983order,lawley2020universal, lawley2020probabilistic, meerson2015mortality}. Here,
we have focused on experimentally relevant $\mathcal{N}$ and extended the theory to account for discrete time, the
calculation of the splitting probability and also the possibility of a heterogeneous population comprised of walkers
with different movement characteristics. As well as addressing the biologically motivated examples, we have utilised the
theory to study some theoretical scenarios on the square lattice. We have uncovered (Fig. \ref{fig: thoer_ffpt})
oscillatory splitting probabilities when persistence is introduced into the population and also shown that the number of
persistent walkers in the population impacts the likelihood that each target in a multi-target search is found before
the other. We have confirmed, in Fig. \ref{fig: thoer_ffpt}(b), that similarly to the one-walker case
\cite{marris2024persistent,tejedor2012optimizing} setting the persistence length too high in a multi-walker search, may
cause an increase in the MfFPT (Fig. \ref{fig: thoer_ffpt}(b)). 

To explain this aspect, consider that the MFPT may be thought of as a measure of the effectiveness of the direct
trajectories \cite{godec2015optimization}. With more effective walkers in the population, the chance of a direct
trajectory being successful increases, which causes the MFPT to fall at a lower time as the mean becomes less skewed by
long meandering trajectories. As a result, the fFPT distribution narrows and becomes more symmetric around the mode. If,
on the other hand, a large subset of the population are ineffective, such as those that are highly persistent and do not
explore the space efficiently, the tails of the distribution will be very long and the MfFPT will increase. In such
extreme cases where the persistence length is much larger than the domain size, we show that replacing diffusive walkers
with persistent walkers in the population actually decreases search efficiency. 

We note that while the mean (and global mean) first-passage times \cite{tejedor2012optimizing,benichou2014first,
condamin2005first, condamin2007first} can be useful in quantifying the overall efficiency of searches, more recent
studies have suggested that the MFPT can often over-estimate the most-probable value for the first-passage time
\cite{mattos2012first} and often falls in the region associated with the indirect trajectories
\cite{godec2015optimization}. Here, the results of Fig. \ref{fig: exp_v_theor} show clearly that the experimental times
in which the food is found are associated with the direct trajectories and are more reminiscent of the modal timescale
than the mean of the fFPT distribution. This corroborates with what we see experimentally where trajectories in the
exploration phase tend to be direct (Fig. \ref{fig: indiv_tracks}). As exemplified by these trajectories, the ants
perform foraging bouts, with looping trajectories of different lengths, which do not meander and explore the entire
space. Instead, these trajectories extend from the nest outwards, and then, after an unsuccessful forage, return to the
nest. 

These spatio-temporal patterns may explain why two of the nine theoretical predictions (exps. 2 and 6) in Fig. \ref{fig:
exp_v_theor} underestimate the empirical findings and show the empirical time falling in a region of near zero
probability of a first first-passage event. After an unsuccessful foraging bout the ant may return to, and remain in,
the nest before performing another foraging excursion. Experimentally, the clock is not stopped if this occurs. This
behavioural trait is not captured by the model, which, by construction, assumes that once a trajectory is started the
walker will search indefinitely in the arena until they find the target site. Therefore, a natural extension to the
model would be to include a mechanism for preferential returns to the initial condition, which may be included via a
discrete space equivalent to the Smoluchowski-type diffusion equation \cite{kac1947random} or via the inclusion of a
process of resetting, whose discrete formulation has been recently derived \cite{das2022discrete,barbinigiuggioli2024}. 

It is known that after the first food source is found, not only do interactions between ants increase, but the number of
ants in the arena also increases \cite{cristin2024spatiotemporal}. More specifically, it was observed that during the
period between the first and second food source being found the number of ants in the system increased monotonically
\cite{cristin2024spatiotemporal}. Recognising a need to incorporate a varying number of ants into the fFPT calculations
when studying foraging strategies, future work may also include the introduction of a dynamic population size into the
fFPT, which is currently known only for continuous one-dimensional diffusion \cite{campos2024dynamic}. Furthermore, the
calculation of the fFPT to the second food item assumes the excursion from the nest to the first found patch is
independent of the excursion from the first patch to the second (or back to the nest), an assumption which is clearly
violated in practice. This highlights the need to develop a theory on \textit{sequential} first-passage times, that is
the first-passage probability to a target $\bm{s}_2$ conditioned on the trajectory first travelling through a previous
target $\bm{s}_1$.

In summary, we have presented a theoretical model to quantify the foraging behaviour of the ant species
\textit{Aphaenogaster senilis} in laboratory experiments. The model extends known theory on correlated random walks to
account for the underlying Y-maze (honeycomb) lattice and on the fFPT of $\mathcal{N}$ independent random walkers,
allowing the population to be heterogeneous in behaviour. By identifying microscopic differences between the ant types,
we have shown that scouts outperform recruits in both search and exploration. We have also highlight that the
characteristic fFPT time of the foraging ants corresponds to the timescale related to the mode of the fFPT distribution
rather than its mean. 

\begin{acknowledgement}
DM acknowledges useful conversations with M{\'a}rton Bal{\'a}zs and funding from an EPSRC DTP studentship, grant number
EP/T517872/1. LG, DM and FB thank the Isaac Newton Institute for Mathematical Sciences, Cambridge, for support and
hospitality during the programmes Mathematics of Movement: an interdisciplinary approach to mutual challenges in animal
ecology and Cell Biology and Stochastic Systems for Anomalous Diffusion, where work was, respectively, initiated and
concluded supported by EPSRC grants EP/R014604/1 and EP/Z000580/1. LG also acknowlegdes funding from NERC grant
NE/W00545X/1 and support by a grant from the Simons Foundation. FB and PF also acknowledge the funding from the Ministry
of Science, Innovation, and Universities in Spain through grant no. PID2021-122893NB-C21.
\end{acknowledgement}
\section*{Appendix 1: Definition of Matrices in Original Master Equation}\label{sec:
trans_mat1}%\addcontentsline{toc}{section}{Appendix 1}
The matrices in Eq. (\ref{eq: vec_ME}) are defined as follows:
\begin{equation}
    \begin{aligned}
        \mathbb{A} &= \{\mathdutchcal{a}_{i,j}\} \text{ where } \mathdutchcal{a}_{3,4}=l, \; \mathdutchcal{a}_{3,5}=r, \;\mathdutchcal{a}_{3,6}=b, \\ 
        \mathbb{B} &= \{\mathdutchcal{b}_{i,j}\} \text{ where } \mathdutchcal{b}_{6,1} = l,\; \mathdutchcal{b}_{6,2}=r, \;\mathdutchcal{b}_{6,3}=b, \\ 
        \mathbb{C} &= \{\mathdutchcal{c}_{i,j}\} \text{ where } \mathdutchcal{c}_{5,1}=r,\; \mathdutchcal{c}_{5,2}=b,\; \mathdutchcal{c}_{5,3}=l, \\ 
        \mathbb{D} &= \{\mathdutchcal{d}_{i,j}\} \text{ where } \mathdutchcal{d}_{2,4}=r,\; \mathdutchcal{d}_{2,5}=b, \;\mathdutchcal{d}_{2,6}=l, \\
        \mathbb{E} & = \{\mathdutchcal{e}_{i,j}\} + c\mathbb{I} \text{ where } \mathdutchcal{e}_{4, 1}=\mathdutchcal{e}_{1, 4}=b, \;\mathdutchcal{e}_{4, 2}=\mathdutchcal{e}_{1, 5}=l,\; \mathdutchcal{e}_{4, 3}=\mathdutchcal{e}_{1, 6} = r, 
    \end{aligned}
\end{equation}
where $\mathbb{I}$ is the $6 \times 6$ identity matrix and where all other indices are zero.

\section*{Appendix 2: The
$z$-Inversion of the Mean Squared Displacement}\label{sec: msd_app}%\addcontentsline{toc}{section}{Appendix 2}

In the diffusive limit Eq. (\ref{eq: MSD_z}) of the main text reduces to 
\begin{equation}
    \widetilde{\Delta\bm{n}}(z) = \frac{2z (c-1) \left[z(4c-1)- 3\right]}{9 (z-1)^2 [(2 c-1) z-1]},
    \label{eq: diff_MSD_z}
\end{equation}
which we invert to the time domain by evaluating the counterclockwise contour integral $\Delta\bm{n}(t) = 1/(2\pi i
)\oint_{|z|< 1}z^{-t-1}\widetilde{\Delta\bm{n}}(z)\text{d}z$ via Cauchy's residue theorem. 

One can easily see the poles of $\widetilde{\Delta\bm{n}}(z)$ lie at $z=1$ (second order) and at $z=1/(1-2c)$ (first
order). Since both these poles are outside the contour region we make the change of variable to $s = 1/z$, such that the
transform becomes $\Delta\bm{n}(t) = 1/(2\pi i )\oint_{|s|\leq 1}s^{t-1}\widetilde{\Delta\bm{n}}(s)\text{d}s$ with the
contour running counterclockwise. We also treat the $c=1/2$ separately.

Starting with the $c=1/2$ case, $\Delta\bm{n}(t) = \frac{1}{18\pi i}\oint_{|s|\leq 1}s^{t-1}(3s-1)(s-1)^{-2}\text{d}s$,
which is trivially evaluated to $\Delta\bm{n}(t) = 9^{-1}(2t+1)$. With $c \neq 1/2$, via Cauchy's residue theorem we
find $\Delta \bm{n}(t) =  \left[4(1-c)t + 1 - (2c-1)^t\right]$. This procedure is equivalent for all cases presented in
the main text.

\section*{Appendix 3: Construction of the Transition Matrices}\label{sec: transition_mats}
%\addcontentsline{toc}{section}{Appendix 3}

When constructing the transition matrices, before resizing them to size $N_1N_2\times N_1N_2$, it is instructive to
consider them as $N_1\times N_2 \times N_1 \times N_2$ sized tensors where the probability at index
$(\alpha,\beta,\gamma,\delta)$ governs transitions from site $\bm{n} = (\gamma,\delta)$ to $\bm{n} = (\alpha, \beta)$.
Therefore, the probability of each movement, which raises (lowers) the $n_1$ coordinate is located on the
$(\gamma+1,\beta,\gamma,\delta)$ $((\gamma-1,\beta,\gamma,\delta))$ diagonal, while those that raise (lower) the $n_2$
coordinate may be found on the $(\alpha,\delta+1,\gamma,\delta)$ $((\alpha,\delta-1,\gamma,\delta))$ diagonal. Sojourn
probabilities are placed on the $(\alpha, \alpha , \alpha, \alpha)$ elements. 

The matrices are sparse as there is one unique way to move between each pair of states. For example, as seen from Eqs.
(\ref{eq: u_ME}) and (\ref{eq: sf}) of the main text, the walker may never leave a positive (negative) site and find
itself in a positive (negative) site in the subsequent time step when it is situated in the bulk of the domain. Hence, the 12
matrices that govern transitions from positive (negative) to positive (negative) sites contain only boundary interactions.
Moreover, the six matrices governing the movement from states $i$ to $i$ ($\forall i \in \{1, 6\}$) are identity
matrices, with the sojourn probability $c$ on the main diagonal, with perturbations at the boundary elements. Finally,
we have 18 matrices governing bulk movement, which each dictate one movement option.

Reflections are included by perturbing the transition matrices at the boundary elements and we model `bouncing off the
wall' as the walker changing its internal state at the boundary site. More specifically, the walker moves to the
internal state that corresponds to the direction that the walker would have been travelling if it had physically hit the
wall and turned around. 

We elucidate this using the upper $\bm{n} = (n_1, N_2)$ edge as an example. On this boundary, it is only the positive
site that interacts with the boundary. If a bounce occurs, on the $\mathbb{Z}^2$ embedding this is akin to the walker
heading North hitting the wall, turning and heading South. This results in a re-wiring of the left and right turns,
which take the walker North out of states 1 and 3, back into state 2 of the same site. Backtracks out of states 1 and 3
are considered as sojourns and re-wired into themselves. The implementation of this boundary condition leads to a
uniform steady state at long times \cite{marris2024persistent}.

Finally, we include absorbing targets by multiplying the outgoing probability of the trap site by $(1-\rho)$, where $\rho$ is the
probability of absorption. More concretely, a trap in a positive state in site $\bm{n}=(n_1,n_2)$ would be included by a
multiplication of $(1-\rho)$ to the $(\gamma,\delta) = (n_1, n_2)$ element of each $\mathbb{M}(i,j)$ $(3 \geq i \geq
6)$.   

\section*{Appendix 4: Experimentally Relevant Parameters}\label{sec: params}
\begin{table}[H]
    \centering
\begin{tabular}{|c|c|c|}
\hline
    \textbf{Parameter:} & \textbf{Meaning:} & \textbf{Value:} \\ \hline
    % Row 1
    $N_1$ & Number of squares in the $x$-direction & 23 \\ 
    % Row 2
    $N_2$ & Number of squares in the $y$-direction & 14 \\ 
    % Row 3
    $(n_{01}, n_{02}, m_0)$ & Initial Condition (Location of Nest) & $(12, 1, \mathfrak{p})$ \\ 
    % Row 3
    $(s_{1}, s_{2}, m_{s})$ & Locations of targets in experiment one. & $(15, 8, \mathfrak{p}), (16, 10, \mathfrak{n})$
    \\ 
    % Row 4
    --- & Locations of targets in experiment two. & $(2, 10, \mathfrak{n}), (19, 12, \mathfrak{n})$ \\ 
    % Row 5
    --- & Locations of targets in experiment three. & $(9, 11, \mathfrak{n}), (10, 12, \mathfrak{p})$ \\ 
    % Row 3
    --- & Locations of targets in experiment four. & $(3, 8, \mathfrak{p}), (16, 4, \mathfrak{p})$ \\ 
    % Row 4
    --- & Locations of targets in experiment five. & $(8, 3, \mathfrak{p}), (15, 8, \mathfrak{p})$ \\ 
    % Row 5
    --- & Locations of targets in experiment six. & $(1, 10, \mathfrak{n}), (12, 5, \mathfrak{p})$ \\ 
    % Row 3
    --- & Locations of targets in experiment seven. & $(7, 5, \mathfrak{p}), (15, 5, \mathfrak{p})$ \\ 
    % Row 4
    --- & Locations of targets in experiment eight. & $(18, 12, \mathfrak{p}), (17, 13, \mathfrak{p})$ \\ 
    % Row 5
    --- & Locations of targets in experiment nine. & $(13, 14, \mathfrak{p}), (17, 8, \mathfrak{n})$ \\ 
    \hline
\end{tabular}
\caption{We use the symbol "---" to denote that the entry is equivalent to the entry above. The two coordinates in the value of the target coordinates refer to the first-found and second-found locations, respectively. These locations are found by matching the distance from the initial condition used in the experimental setup.}
\end{table}

\end{document}